%% file: main.tex
\documentclass[a4paper,11pt]{article}
\pdfoutput=1 

\newcommand{\degC}{$^{\circ}$C~}
\newcommand{\neut}{$\rm n_{eq}/cm^2$}
\newcommand{\vendorH}{HPK}
\newcommand{\vendorF}{FBK}
\newcommand{\vendorOne}{STP}
\newcommand{\vendorFive}{JTC}

\usepackage{orcidlink}
\usepackage{lineno}
\usepackage{setspace}
\usepackage{verbatim}
\usepackage{amsmath}

\usepackage{standalone}
\usepackage{siunitx}
\usepackage{import}
\usepackage{tikz}
\usepackage{graphicx}
\usepackage{subcaption}

\usepackage[version=4]{mhchem}

\usepackage{multirow}

\usepackage{jinstpub}
\usepackage{cleveref}
\title{Optimization of LYSO crystals and SiPM parameters for the CMS MIP timing detector}

\newcommand\dnote[2][]{%
\setcounter{footnote}{1}
\renewcommand{\thefootnote}{\fnsymbol{footnote}}
\if!#1!%
\stepcounter{footnote}\footnotetext{#2}%
\else%
{\renewcommand\thefootnote{#1}%
\footnotetext{#2}}%
\fi}

\input{authors.tex}

\input{address.tex}

\emailAdd{flavia.cetorelli@cern.ch}
\emailAdd{simona.palluotto@cern.ch}

\abstract{

For the High-Luminosity (HL-LHC) phase, the upgrade of the Compact Muon Solenoid (CMS) experiment at CERN will include a novel MIP Timing Detector (MTD). 
The central part of MTD, the barrel timing layer (BTL), is designed to provide a measurement of the time of arrival of charged particles with a precision of 30~ps at the beginning of HL-LHC, progressively degrading to 60~ps while operating in an extremely harsh radiation environment for over a decade.
In this paper we present a comparative analysis of the time resolution of BTL module prototypes made of LYSO:Ce crystal bars read out by silicon photo-multipliers (SiPMs). 
The timing performance measured in beam test campaigns is presented for prototypes with different construction and operation parameters, such as different SiPM cell sizes (15, 20, 25 and 30~$\rm \mu m$), SiPM manufacturers and crystal bar thicknesses. The evolution of time resolution as a function of the irradiation level has been studied using non-irradiated SiPMs as well as SiPMs exposed up to $2\times 10^{14}$~\neut~ fluence. 
The key parameters defining the module time resolution such as SiPM characteristics (gain, photon detection efficiency, radiation induced dark count rate) and crystal properties (light output and dimensions) are discussed.
These results have informed the final choice of the MTD barrel sensor configuration and offer a unique starting point for the design of future large-area scintillator-based timing detectors in either low or high radiation environments.
}


\begin{document}

\maketitle

\flushbottom
\section{Introduction}

The upgrade program of the Compact Muon Solenoid experiment for the High-Luminosity phase of the Large Hadron Collider (HL-LHC) at CERN includes the integration of a new detector, the MIP Timing Detector (MTD), designed to measure the time of arrival of charged particles with a precision of 30 ps at the beginning of HL-LHC operations, progressively degrading to 60 ps after irradiation. 
The track-time information for charged particles will help disentangle the approximately 200 concurrent collision vertices per beam crossing (pileup events) expected at the HL-LHC. A resolution between three and six times better than the time spread of the luminous region (200~ps RMS) will recover the quality of the event reconstruction of current LHC pileup levels ~\cite{CMS_MTD_TDR}. Moreover, time-of-flight information will bring new capabilities to the CMS detector by expanding the physics reach in searches for long-lived unstable particles, providing particle identification for heavy-ion collisions, and improving exclusive decay reconstruction in heavy flavour physics ~\cite{CMS_MTD_TDR, CMS-DP-2022-025}.

The central part of the MTD, the Barrel Timing Layer (BTL), is a thin cylindrical layer at a radial distance of 1 m from the beam axis, covering the pseudorapidity region $|\eta| < 1.5$. The basic detection unit of the BTL modular structure consists of an array of 16 elongated LYSO:Ce scintillating crystal bars, with approximate dimensions of 55 $\times$ 3 $\times$ 3~mm$^3$, coupled to arrays of 16 silicon photomultipliers (SiPMs) at the two ends. The ratio of the SiPM to the crystal surface is optimized to provide the required performance over a large surface (38~m$^2$) within power, channel count, services, and cost constraints. This configuration was proven to give a resolution better than 30 ps with single LYSO:Ce bars coupled to non-irradiated SiPMs~\cite{BTL_TB_paper_2021}.

One of the major challenges of the BTL operation is the increase of leakage current in the SiPMs after irradiation.
A dark count rate (DCR) of up to a few tens of GHz per SiPM is foreseen for the integrated radiation levels expected after 3000~fb$^{-1}$ (corresponding to a fluence of $2\times10^{14}$ \neut~ and to a total ionizing dose of 32~kGy), and represents the dominant contribution to the time resolution at the end of operation. 
To our knowledge, no other experiment has ever envisaged to use SiPMs in such a harsh radiation environment. 
Key innovations to counter the SiPM DCR noise include the first in-silicon implementation of the differential leading edge discrimination concept~\cite{DLED_GolaPiemonte} in the TOFHIR2 readout ASIC~\cite{TOFHIR2_Proceeding, Albuquerque_2024} and the smart thermal management of SiPMs, with operation at T$_{\rm op}$~=~-45$^{\circ}$C and in-situ annealing at T$_{\rm a}$~=~+60$^{\circ}$C during the HL-LHC stops~\cite{Bornheim_2023}.

In this work, we describe the optimization of the SiPM parameters and of the BTL module geometry, with the comparative analysis of several LYSO:Ce arrays assembled both with new SiPMs and with SiPMs irradiated up to the total radiation level expected at the end of the BTL operation. Section~\ref{sec:drivers} introduces the time resolution dependence on key sensors parameters. The sensor modules tested and the experimental setup are described in section~\ref{sec:modules} and section~\ref{sec:experimentalSetup}, while details on the analysis methods are given in section~\ref{sec:analysis}.
Comparisons of the timing performance achieved for modules with SiPMs of various cell sizes (15, 20, 25 and 30~$\mu$m), with different crystal thickness and with SiPMs irradiated at different neutron fluences are discussed in sections ~\ref{sec:cell-sizes}, ~\ref{sec:geometries} and ~\ref{sec:fluences}.

The current study describes the optimization of the crucial parameters for scintillator and SiPM-based large area timing detectors in order to reach a resolution of a few tens of ps in either low or high radiation environments.
The results demonstrate that a time resolution of about 25~ps at the beginning of operation and about 60~ps at the end of operation are achieved with 3.75~mm thick crystal arrays coupled to 25~$\rm \mu$m cell size SiPMs, which has thus been identified as the final sensor configuration for the BTL detector.

\section{Time resolution drivers}
\label{sec:drivers} 

The BTL detector provides a single time measurement per track from the average of the time measurements at the two ends of the bars. Leading edge discrimination and amplitude measurements of the SiPM pulses from the scintillation light are performed by the TOFHIR ASICs~\cite{Albuquerque_2024}.

The time resolution of the BTL detector is determined by different factors and can be expressed as the sum in quadrature of various contributions:

\begin{equation}
\sigma_t = \sigma_t^{ele} \oplus
    \sigma_t^{phot} \oplus 
    \sigma_t^{DCR} \oplus 
    \sigma_t^{clock}
\end{equation}

where $\sigma_t^{ele}$ is the contribution from the electronic noise, $\sigma_t^{phot}$  from the photo-statistics, $\sigma_t^{DCR}$ from the DCR noise,
and $\sigma_t^{clock}$ from instabilities of the clock distribution.
Dedicated studies showed that clock instabilities are within the design specifications ($<$~10~ps)~\cite{SIMKINA_proceeding}. 
The other three terms are comparable in size and the sensor’s R\&D effort has been directed towards their optimization.

The time jitter due to the electronic noise depends on the steepness (dI/dt) of the signal pulses at the leading edge discrimination threshold and can be expressed as

\begin{equation} 
\sigma_t^{ele} = \frac{\sigma_{noise}}{dI/dt} \oplus \sigma^{TDC}_t
\label{eq:ele}
\end{equation}

where $dI/dt$ is the signal slope at the discriminator input and $\sigma^{TDC}_t$ is the contribution from the TDC. From laboratory measurements with blue laser light directly illuminating naked SiPMs, we estimated $\sigma_{noise} = 420 \pm 35$~nA and $\sigma^{TDC}_t = 16 \pm 2$~ps.
The design of the FE analog part (bandwidth and series impedance) was optimized independently, as discussed in~\cite{Albuquerque_2024}. In this comparative study,  we focus only on the effects on the dI/dt related to the SiPM: the rising slope varies depending on the specific SiPM choice and operating over-voltage \footnote{excess bias beyond the break-down voltage} and to first approximation scales as $\propto N_{pe} \cdot G \cdot f_{SiPM}$ where $N_{pe}$ is the number of photo-electrons produced, G is the SiPM gain and $f_{SiPM}$ is a factor related to the electrical properties of the SiPM (quenching resistor, cell capacitance, grid capacitance, etc...). 

The photo-statistics term, due to the statistical fluctuations in the time of arrival of the photons detected at the SiPMs, scales approximately as

\begin{equation} 
\sigma_t^{phot} \propto \frac{1}{N_{pe}^{  \alpha}}  ~~~~\rm ~~~~~~~~~ with ~~\alpha\approx0.5
\label{eq:phot}
\end{equation}
The DCR term increases with the dark count rate in the SiPMs approximately as

\begin{equation} 
\sigma_t^{DCR} \propto \frac{DCR^{\beta}}{N_{pe}} ~~~~\rm ~~~~~ with ~~\beta\approx0.5
\label{eq:dcr}
\end{equation}
From eq.~\ref{eq:ele}, \ref{eq:phot}, \ref{eq:dcr}, it is evident that a key parameter to optimize the time resolution is the number of photo-electrons which depends on a variety of factors as:
\begin{equation}\label{eq:nphe}
    N_{pe} = E_{dep}\cdot LY \cdot LCE \cdot PDE
\end{equation}
where $E_{dep}$ is the amount of ionization energy deposited in the crystal by the charged particle, $LY$ the crystal light yield (number of photons produced per MeV of energy deposit), the light collection efficiency, $LCE$, is the fraction of photons reaching the SiPM window (determined by the crystal dimensions, wrapping and optical couplings~\cite{Addesa_2022}) and the PDE is the SiPM photon detection efficiency.

\begin{figure}[!tbp]
    \centering
    \includegraphics[width=0.49\linewidth]{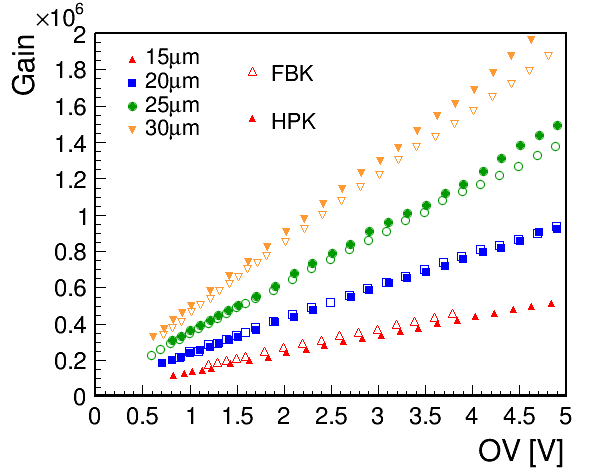}
    \includegraphics[width=0.49\linewidth]{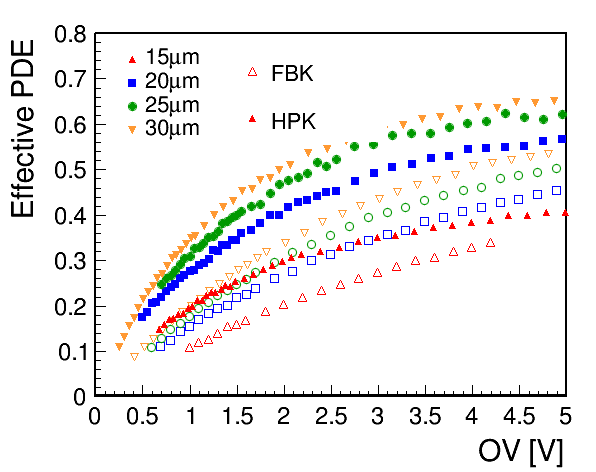}
    \caption{Gain (left) and effective PDE (right) as a function of over-voltage for SiPMs with different cell size and from different manufacturers. The effective PDE corresponds to the measured $PDE(\lambda)$ convoluted with the emission spectrum of LYSO:Ce crystals, which peaks around 420~nm.}
    \label{fig:sipm_gain_pde}
\end{figure}

Increasing the SiPM cell size is one possible way to optimize the time resolution as both the SiPM gain and PDE increase with larger cells as shown in figure~\ref{fig:sipm_gain_pde}. The fill factor of the SiPMs increases with cell sizes of 15, 20, 25 and 30~$\mu m$ respectively as 0.61 (0.65), 0.70 (0.73), 0.76 (0.77), 0.80 (0.80) for HPK (FBK). The effective PDE shown in the figure corresponds to the measured PDE as a function of the light wavelength $PDE(\lambda)$ (which is different between \vendorF~ and \vendorH) convoluted with the emission spectrum of LYSO:Ce crystals, which peaks around 420~nm.

In particular, the SiPM gain depends on the SiPM cell capacitance $C_d$ and quenching capacitance $C_q$ as $G = V_{OV} \cdot (C_q + C_d) / e$, with $V_{OV}$ the over-voltage and $e$ the electron charge, and $C_d$ and $C_q$ expected to scale proportionally to the cell area and cell perimeter, respectively.
 The PDE scales with the fill factor which is larger for larger cell-sizes. 
On the other hand, the DCR also depends on the SiPM effective active area and thus increases proportionally to the PDE. However, given that the DCR term in the time resolution varies only as the square root of the DCR (eq.~\ref{eq:dcr}), SiPMs with larger cell-size can still be advantageous for irradiated devices, as will be shown in section~\ref{sec:cell-sizes}.

The adverse effects that play a role when increasing the cell size are related to radiation damage effects. In particular, for the same level of DCR a larger fraction of cells will be busy due to a dark count and blind to incoming photons leading to an effective loss of PDE. This effect which is further amplified by the fact that larger cells have a longer recharge time, $\tau = R_q (C_q+C_d)$, remains below 5\% (relative), for the largest DCR levels measured with $30~\mu m$ SiPM cell size.
Another drawback of SiPMs with a larger gain is the corresponding increase in the detector power consumption and in the amount of heat produced by the SiPM which can lead to SiPM self-heating. 

The time resolution ultimately depends on the number and time distribution of the photo-electrons produced by the interaction of a charged particle in the crystal. After irradiation, this number depends on the crystal thickness, the SiPM properties and the operating conditions in a non-trivial way. For this reason, the choice of the optimal cell size was not obvious and required a thorough characterization with irradiated devices. We studied several BTL modules with SiPMs of cell-size ranging from 15~$\rm \mu$m to 30~$\rm \mu$m, spanning a range where we expected a reasonable number of photo-electrons and DCR levels that could be handled with appropriate cooling and in-situ annealing.

\section{Description of sensor modules tested}\label{sec:modules} 
The thicker sensor module that fits the CMS envelope constraints consists of an array of crystal bars in which each bar has dimensions (length $\times$ width $\times$ thickness) of $55.0~\times~3.12~\times~3.75$~mm$^3$ (T1 module). 
Thinner module prototypes, 3.0 mm (T2) and 2.4 mm (T3) thick, were also studied as an option to instrument regions at medium and high pseudorapidity, maintaining the same slant thickness for particles coming from the interaction point.
Each bar is optically isolated from the neighboring bars by means of 80~$\rm\mu m$ thick reflective foils of Enhanced Specular Reflector (3M ESR). 
The two opposite ends of each bar in the array are attached by means of a 100~$\rm\mu m$ layer of RTV3145 glue to a pair of SiPM arrays.
The SiPM arrays feature a custom design for optimal heat dissipation; they include a RTD temperature sensor, four small thermoelectric coolers (TECs) for temperature control and stabilization and a flex cable to route the signal to a front-end (FE) board~\cite{Bornheim_2023}.
The FE board houses two voltage regulator chips (ALDOs~\cite{ALDO_2024}) to adjust the bias voltage (BV) of a SiPM array (the same BV is provided to all 16 SiPMs inside one array) and a dedicated readout chip (TOFHIR) that can perform charge integration measurement, time-over-threshold (ToT) measurements and features a TDC with 20~ps binning to record the time when a given pulse crosses configurable thresholds~\cite{TOFHIR2_Proceeding, Albuquerque_2024}.

\begin{figure}[!tbp]
    \centering
    \includegraphics[width=0.99\linewidth]{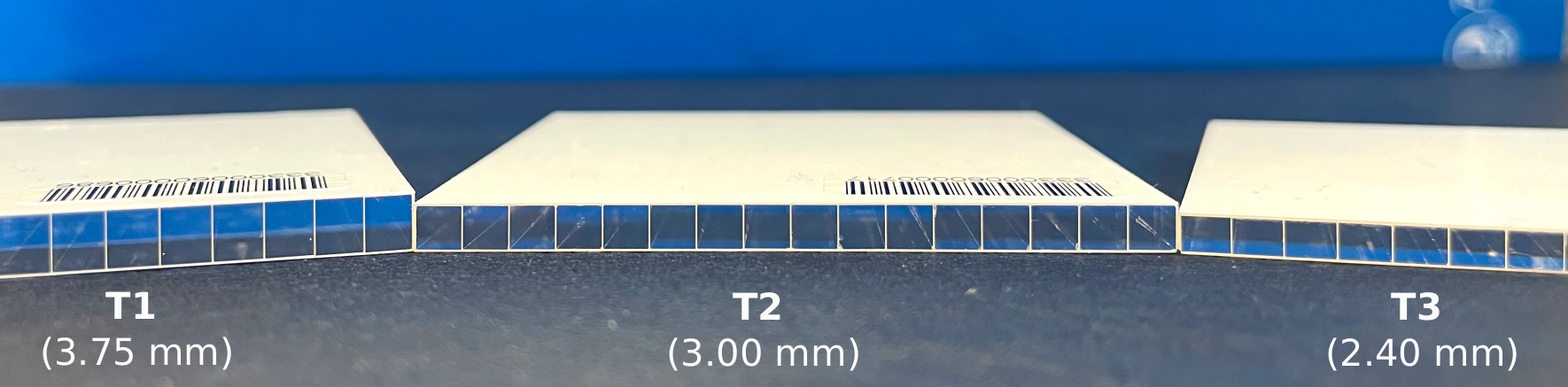}
    \includegraphics[width=0.59\linewidth]{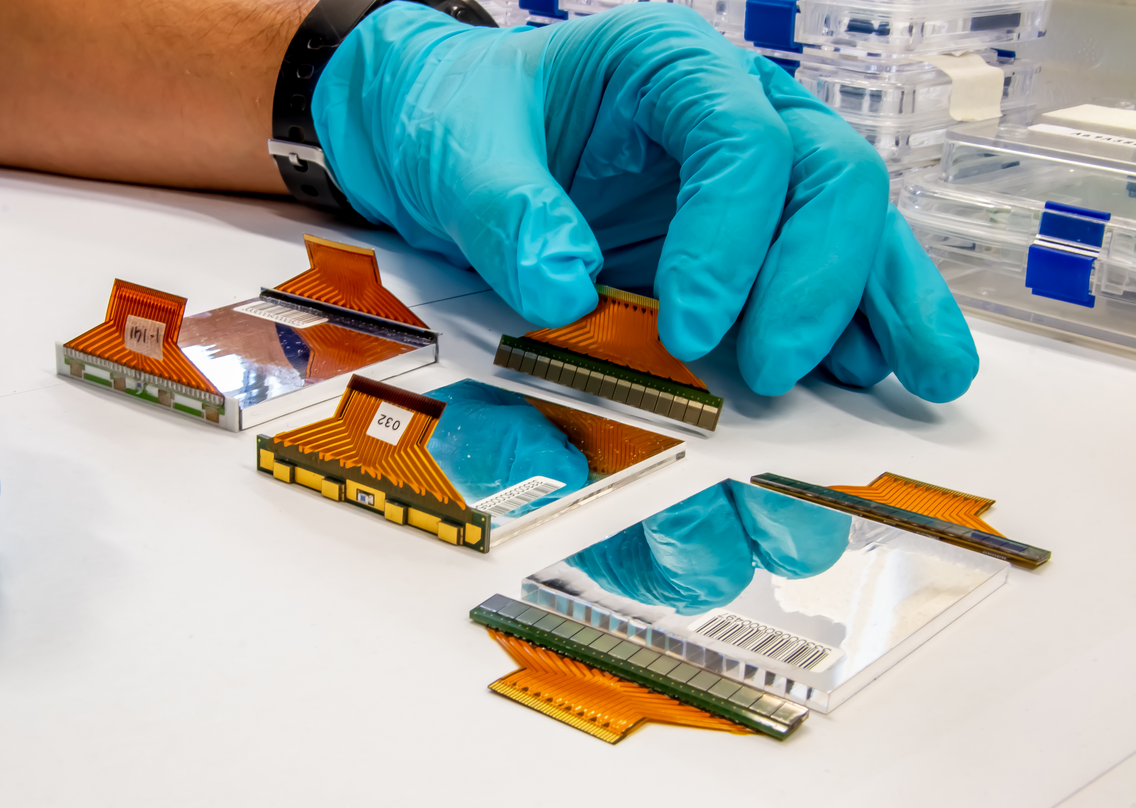}
    \includegraphics[width=0.40\linewidth]{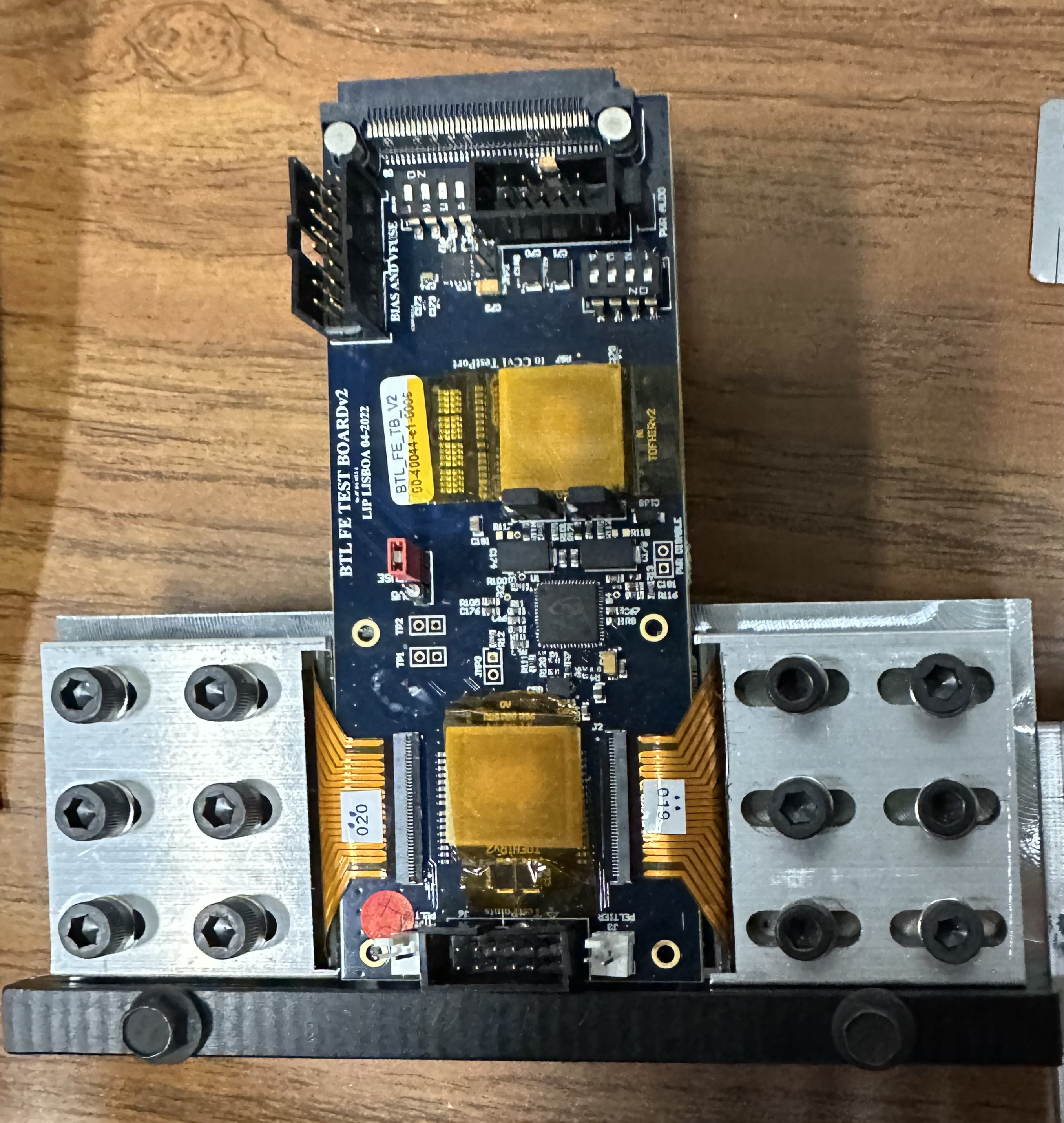}
    \caption{Top: picture of LYSO arrays of T1 (3.75 mm thickness), T2 (3.00 mm thickness) and T3 (2.40 mm thickness) from left to right. Bottom left: picture of three BTL modules after and before gluing of the SiPM arrays to the crystal array. Bottom right: picture of a module with the flex cables connected to the front end board and with the SiPM back side coupled thermally to a aluminum heat sink used during the test.}
    \label{fig:modules_picture}
\end{figure}

The crystals in T1, T2, and T3 modules are coupled to SiPM arrays with an active area matched to the crystal end face for optimal light collection.
The crystal arrays were produced according to common specifications by two manufacturers, Sichuan Tianle Photonics (\vendorOne) and Suzhou JT Crystal Technology (\vendorFive), and demonstrated to have comparable light yield and decay time in previous laboratory tests \cite{Addesa_2022}, thus no difference in the timing performance of the modules is expected due to the different crystal origin.
A picture of three LYSO arrays of different thickness and of a few sensor modules tested on beam is shown in figure~\ref{fig:modules_picture}.

Sensor modules were assembled using SiPM arrays with different cell sizes of 15, 20, 25 and 30~$\rm \mu m$.
The SiPM arrays tested were also produced by different manufacturers: Hamamatsu Photonics (\vendorH) and Fondazione Bruno Kessler (\vendorF). A different silicon wafer technology is used by the two manufacturers yielding different SiPM properties such as PDE, gain, excess charge factor (ECF). SiPMs are also affected differently by radiation damage leading to different levels of DCR, different drifts of breakdown voltage with irradiation and different dependencies of DCR on temperature.
More details on these differences and on the characterization of these SiPMs before and after irradiation can be found in \cite{Musienko_2017, Bornheim_2023}.
The SiPM encapsulation used to protect the SiPM active area from accidental scratches is also different: HPK SiPMs feature a 300~$\rm \mu m$ silicone resin while FBK SiPMs are protected by a 50~$\rm \mu m$ thin layer of sputtered glass yielding a higher efficiency in collecting scintillation light from the crystal. Both coatings were demonstrated to be radiation tolerant up to the maximum radiation levels expected in the BTL detector \cite{CMS_MTD_TDR}, showing less than 5\% loss in transparency.

Some of the modules were assembled using non-irradiated SiPMs to evaluate the performance expected at the beginning of the MTD detector operation while another set of modules was built using irradiated SiPMs to study the degradation of time resolution throughout the entire detector operation period.
The irradiation of the SiPM arrays was performed at the JSI neutron reactor in Ljubjana up to an integrated fluence of $2\times10^{14}$ \neut, corresponding to the radiation level inside the MTD barrel detector after 3000~fb$^{-1}$.

Since radiation damage effects will partially anneal spontaneously at a temperature-dependent rate inside the MTD detector, an accelerated annealing procedure was also performed on the SiPM arrays used for the tested modules to reproduce the expected annealing effects in the real detector. The SiPM arrays underwent an annealing sequence of 40 minutes at 70$^{\circ}$C, three days at 110$^{\circ}$C and four days at 120$^{\circ}$C.
According to the annealing model derived from \cite{MollThesis}, this annealing scheme decreases the DCR by a factor of two more than the annealing foreseen in the BTL detector by increasing the temperature to 60\degC during the accelerator technical stops scheduled over about 10 years of HL-LHC operation.
For this reason, in the test beam studies the SiPMs irradiated to an integrated fluence of $2\times10^{14}$ \neut~ have been operated at a temperature of -35$^{\circ}$C (10\degC higher than the planned BTL operating temperature) to reproduce the DCR conditions expected at the end of the detector operation. 

\begin{figure}[!tbp]
    \centering
    \begin{subfigure}[b]{0.49\textwidth}
        \centering
        \includegraphics[width=\textwidth] {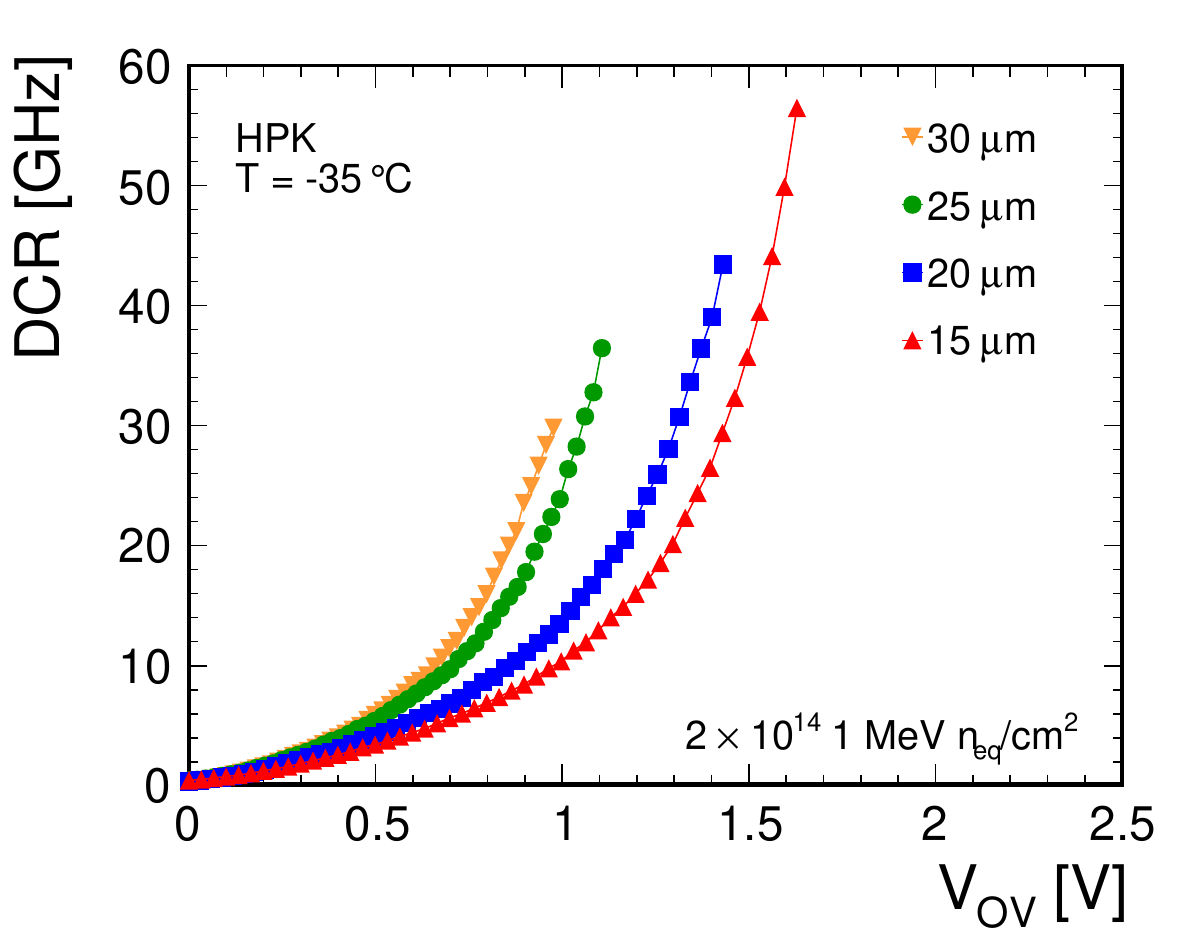}
        \caption{}
    \end{subfigure}
    \hfill  
    \begin{subfigure}[b]{0.49\textwidth}
        \centering
        \includegraphics[width=\textwidth]{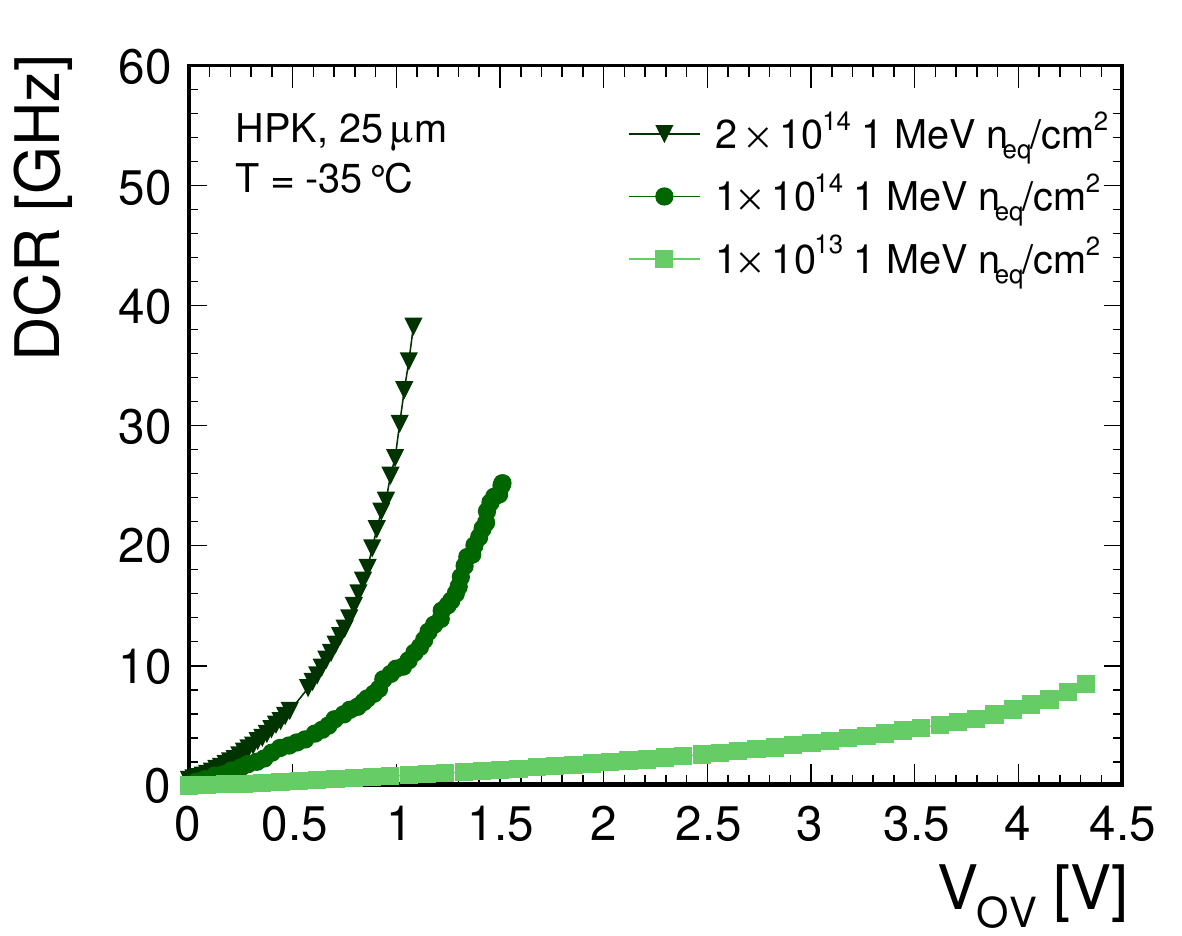}
        \caption{}
    \end{subfigure}
    \hfill
    \begin{subfigure}[b]{0.49\textwidth}
    \centering
    \includegraphics[width=\textwidth]{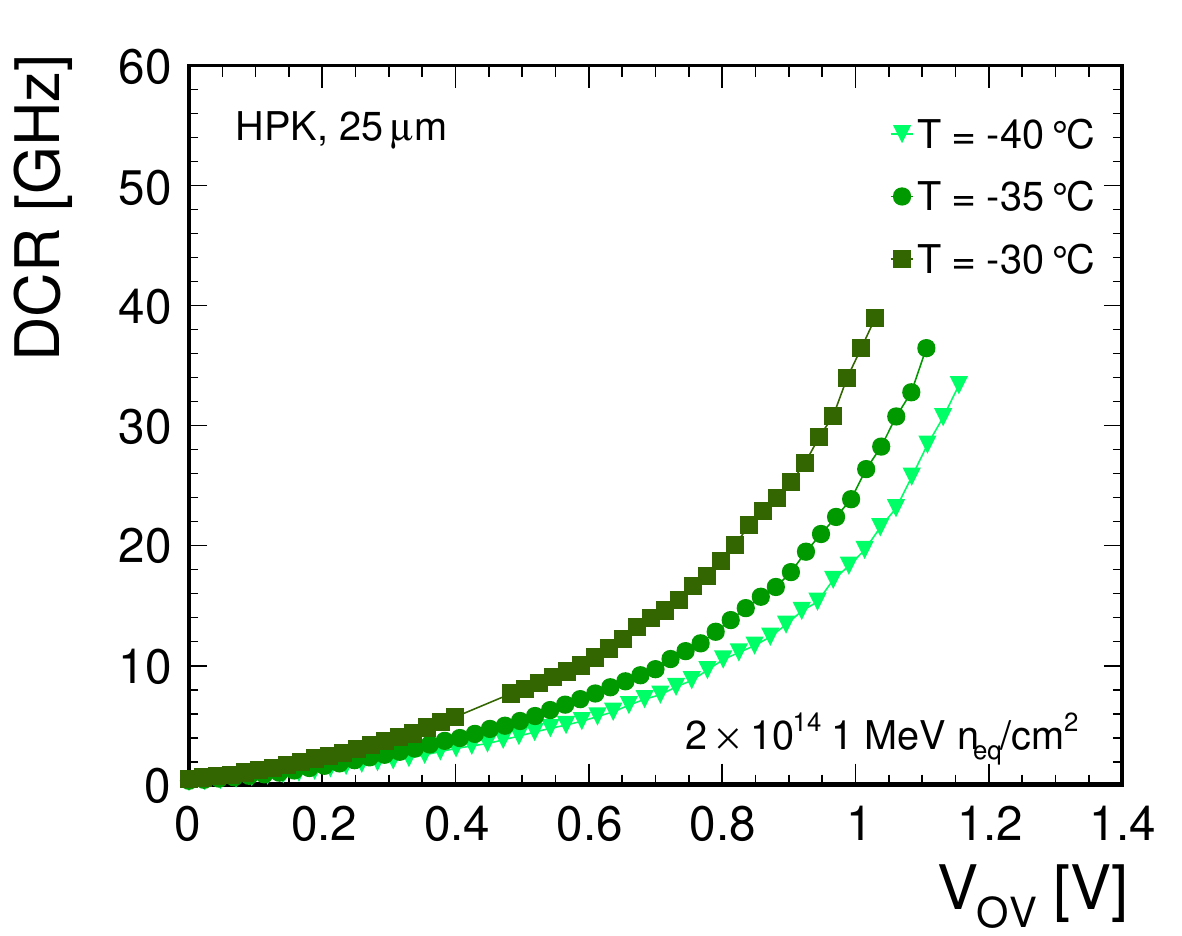}
    \caption{}
    \end{subfigure}
    \hfill
    \begin{subfigure}[b]{0.49\textwidth}
        \centering
        \includegraphics[width=\textwidth]{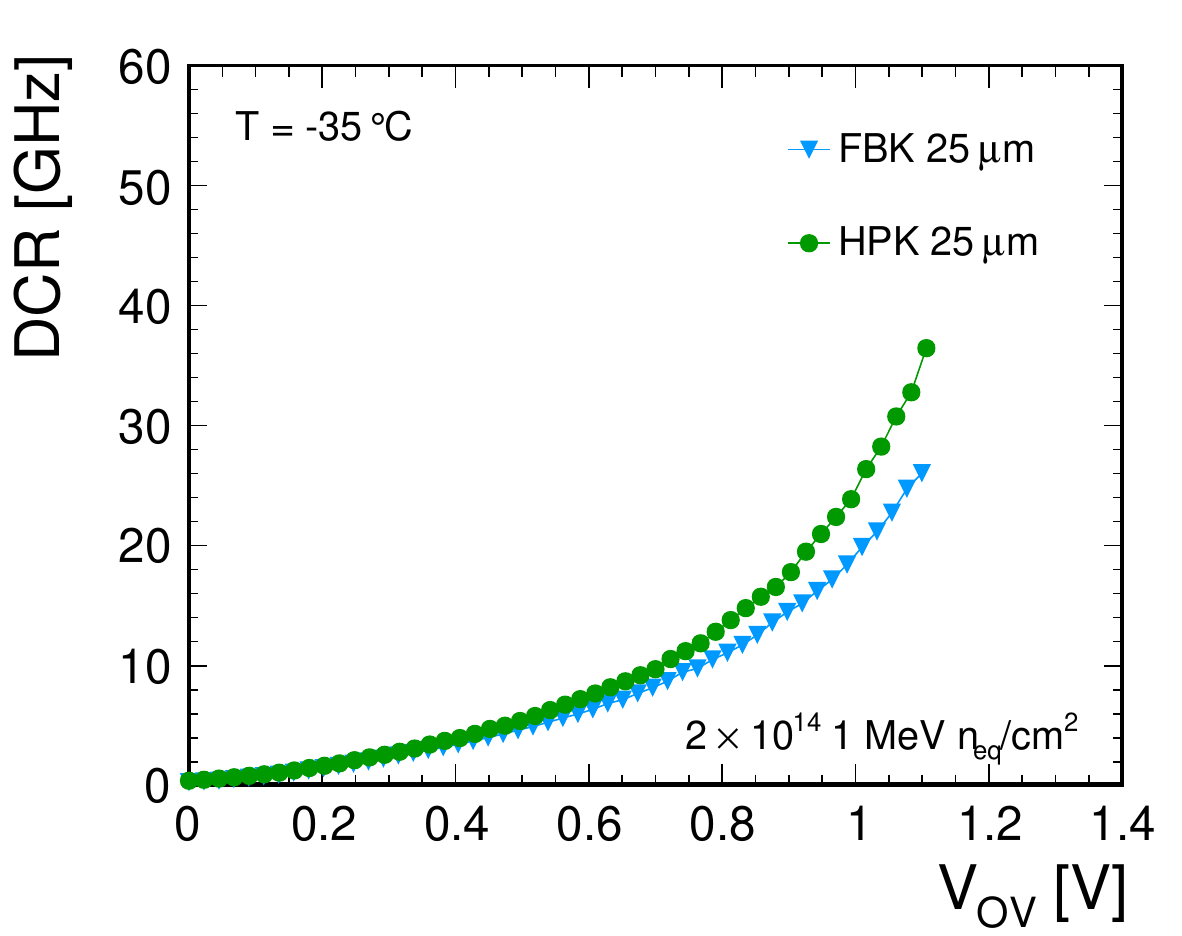}
        \caption{}
    \end{subfigure}
    \caption{Single SiPM dark count rate  as a function of the SiPM over-voltage for different configurations tested: (a) T2 modules with SiPMs of different cell-size (15, 20, 25, 30~$\rm\mu m$) irradiated to $2\times10^{14}$~\neut; (b) T1 modules with 25~$\rm\mu m$ cell-size SiPMs irradiated to different fluences ($1\times10^{13}$, $1\times10^{14}$, $2\times10^{14}$~\neut); (c) T2 module with 25~$\rm\mu m$ cell-size HPK SiPMs irradiated to $2\times10^{14}$~\neut for different operating temperatures; (d) T2 modules with 25~$\rm\mu m$ cell-size SiPMs from HPK and FBK irradiated to $2\times10^{14}$~\neut. The SiPM arrays underwent the same annealing sequence, as described in the text.}
    \label{fig:dcr-curve}
\end{figure}

The DCR of a single SiPM inside an irradiated array is shown for different modules at a temperature of -35\degC in figure~\ref{fig:dcr-curve} as a function of the SiPM over-voltage ($\rm{V_{OV}}$). The DCR is estimated from a measurement of the SiPM array current ($\rm I_{array}$) as 

\begin{equation}
{\rm  DCR} = \frac{\rm I_{array}/16}{\rm G(V_{OV})\cdot e}
\label{eq:dcr_calc}
\end{equation}
where $\rm G(V_{OV})$ is the SiPM gain, e is the electron charge and the array current $\rm I_{array}$ is assumed to be equally distributed among the 16 channels in the SiPM array.

For the same over-voltage, SiPMs with larger cell-size show a larger DCR compared to SiPMs with smaller cells (figure~\ref{fig:dcr-curve}~(a)) due to their higher fill factor and thus an effectively larger active area. A faster than linear growth of the DCR as a function of the dose is observed for large fluences, as shown in figure~\ref{fig:dcr-curve}~(b). 
The DCR also depends on the temperature, 
as reported in figure~\ref{fig:dcr-curve}~(c), increasing by about a factor 2 over 10\degC.
The effect of radiation damage is also slightly different between SiPMs from the two producers, with FBK SiPMs showing about 20\% less DCR than HPK ones for an over-voltage of 1~V, corresponding to a typical end-of-operation working point (figure~\ref{fig:dcr-curve}~(d)).

\section{Experimental setup and procedures}
\label{sec:experimentalSetup}

A test beam campaign was carried out during 2022 and 2023 to evaluate the time resolution of the set of modules described in section~\ref{sec:modules} for minimum ionizing particle (MIP) detection.
Sensor modules were tested using FE test boards with TOFHIR2 ASICs (of either the 2X or 2C version, which were proven to be equivalent in terms of impedance, bandwidth, and the time resolution they provide) \cite{TOFHIR2_Proceeding, Albuquerque_2024}. The FE boards are readout using a FEB-D readout board connected to a PC.

\begin{figure}[!tbp]
    \centering
    \includegraphics[width=0.99\linewidth]{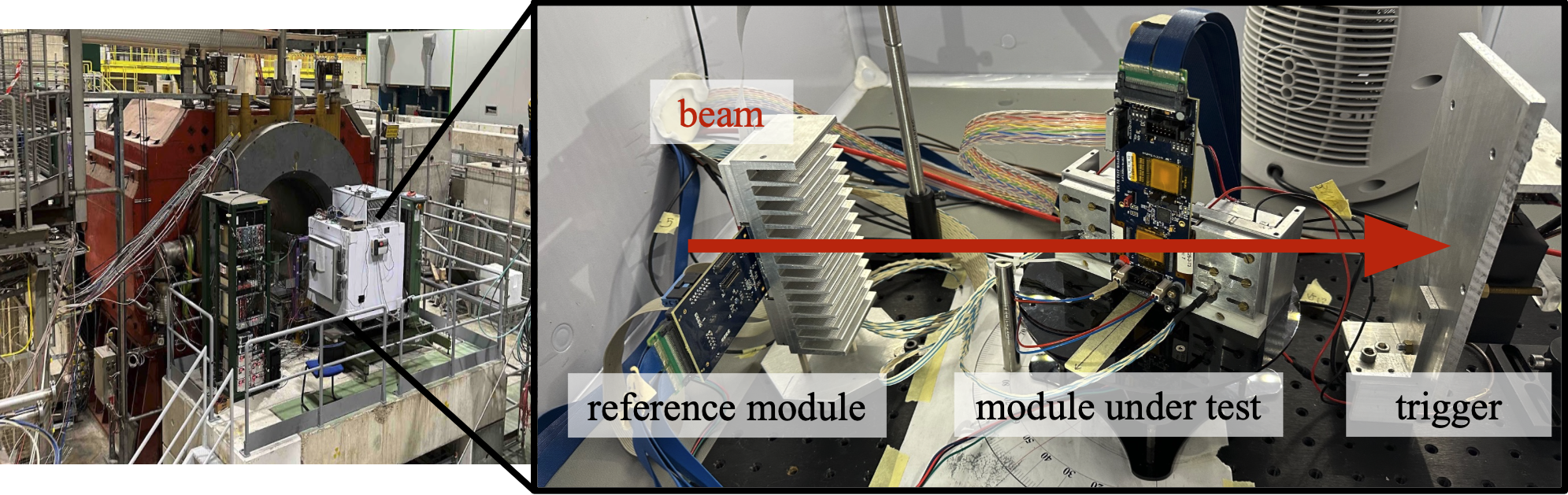}\\
    \includegraphics[width=0.9\linewidth]{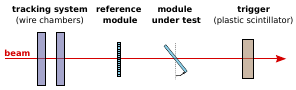}
    \caption{Top: picture of the cold box installed on the H8 beam line at the SPS CERN facility (left) and of the experimental setup inside the box consisting of the reference module, the module under test (right). Bottom: schematic plan view of the experimental setup showing the rotation of the module under test with respect to the beam direction.}
    \label{fig:exp_setup_photo}
\end{figure}

A picture and a schematics of the setup are shown in figure~\ref{fig:exp_setup_photo}.
A tracking device was located upstream along the beam line. Two sensor modules were readout simultaneously; they were positioned with a relative orientation of $90^{\circ}$.
The upstream module was placed at normal incidence angle compared to the beam direction and was used as reference to provide both a precise time stamp ($\sim30$~ps time resolution) and a coarse position information ($\sim3$~mm) of the incoming particles on the horizontal plane. The second module, referred to as device under test (DUT), was tilted by a certain angle $\theta$, compared to the beam direction to emulate the average energy deposit of charged particles expected in the MTD barrel detector due to the spread in particle impact angle on the crystal. 
The angle of the DUT was adjusted using a remotely controlled rotating stage. The angles are measured with a precision of about $1^\circ$, estimated from a scan of the MIP energy most probable value as a function of the angle.
A thick plastic scintillator located downstream of the DUT with dimensions of $7\times7$~cm$^2$ was used to generate an acquisition trigger. 

The sensor modules were placed inside a light tight cold box in which temperature could be regulated in the range between $-$35\degC and 15\degC. The thermoelectric coolers on the SiPM package were used to stabilize (within $\pm~1$\degC) the SiPM operating temperature to 5-10 degrees below the box environmental temperature.
The modules and the respective FE boards were kept in position by metallic supports that also served as heat sinks instrumented with fans to optimally extract the heat produced by the operation of irradiated SiPMs.

Measurements with the same setup were performed both at the Fermilab Test Beam Facility (FTBF) and at the CERN SPS H8 beam line using 120 GeV protons and 180~GeV pions respectively. In either case, the charged particles deposit a comparable amount of energy inside the crystal (according to a Landau distribution with most probable value around 0.86 MeV/mm) yielding the same time resolution for a certain module. Results on the same modules obtained both at CERN and FNAL were demonstrated to be equivalent. 

The TOFHIR2 readout ASIC has three leading-edge current discriminators per channel, two in the timing branch of the chip and one in the energy branch. For the measurements reported in this paper, the chip was configured to use one leading edge discrimination threshold to extract the time measurement on the rising edge of the pulse; the second time measurement is extracted from the crossing of the same threshold on the falling edge of the pulse and is used, together with the first time measurement, to measure the time-over-threshold (ToT). The crossing of the first threshold also defines the start of the charge integration. More details about TOFHIR can be found in \cite{TOFHIR2_Proceeding, Albuquerque_2024}.

\section{Data analysis methods}
\label{sec:analysis}

\begin{figure}[!t]
\begin{center}
\includegraphics[width=0.49\textwidth]{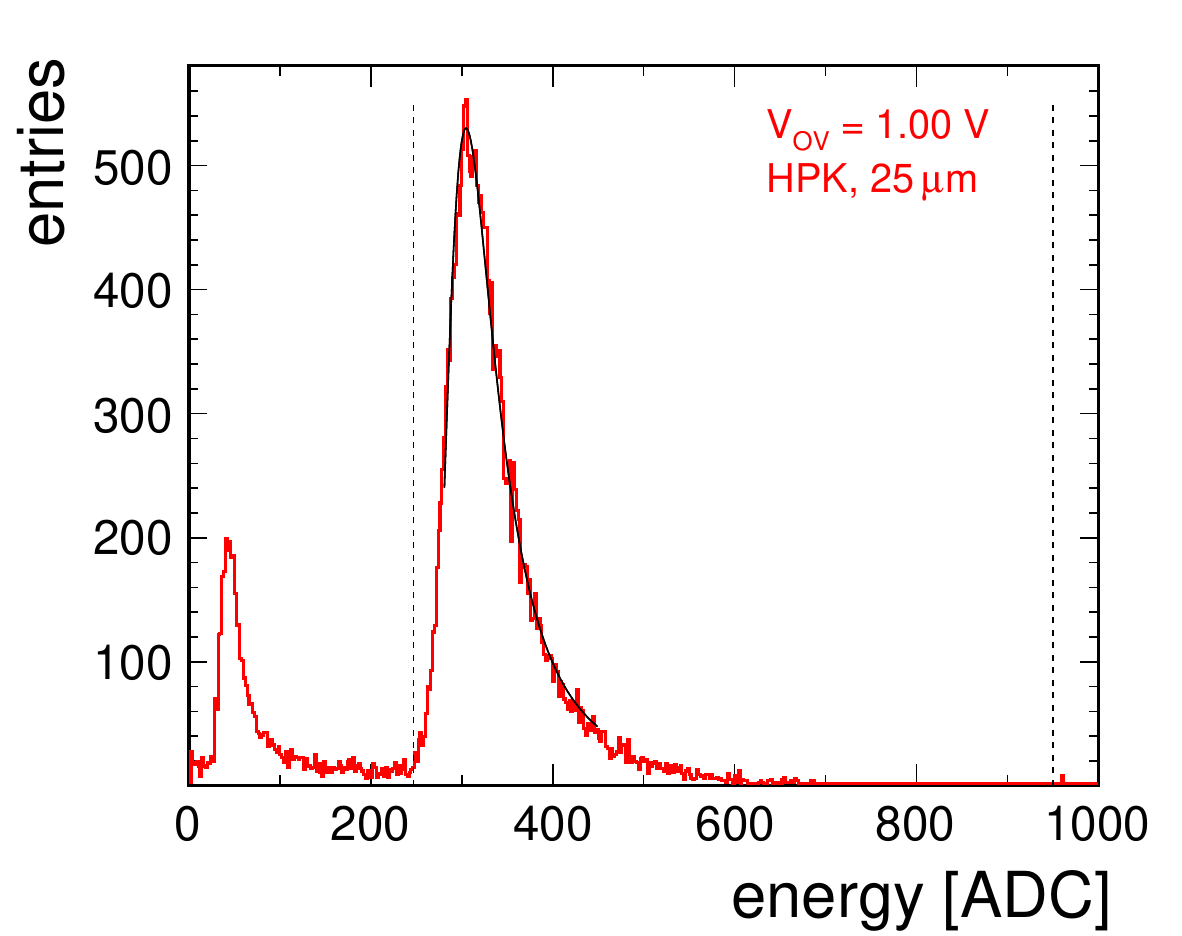}
\includegraphics[width=0.49\textwidth]{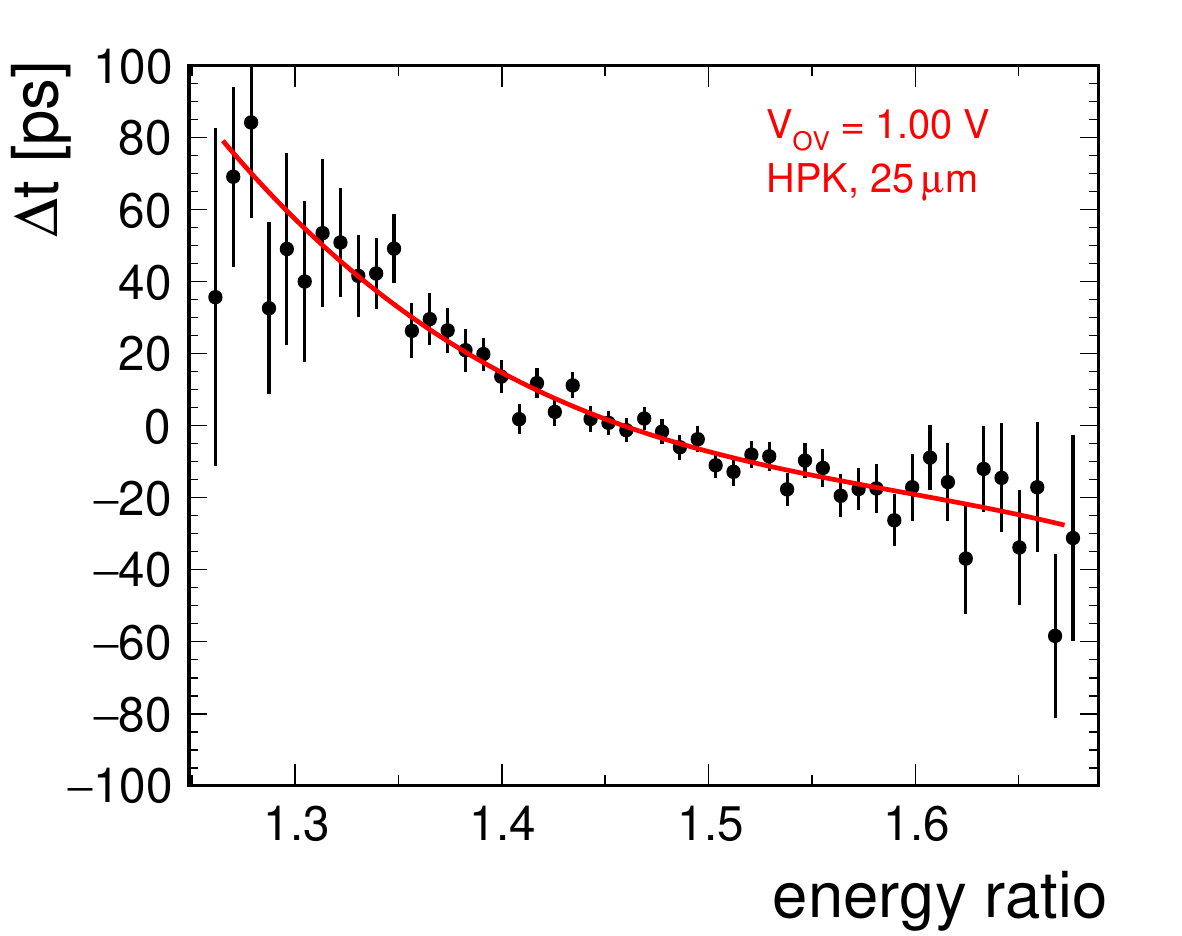}  
\end{center}

\caption{Left: example of MIP energy distribution measured in one bar. The energy is the average between the energies measured by the two SiPMs. The distribution is fitted with a Landau function. Events with energy in the range between the vertical dashed lines are retained for the analysis; events at low energy are due to cross-talk from adjacent bars. Right: example of dependence of $t_{diff}$ on the energy ratio between the two channels. The module with HPK SiPM of 25$~\mu$m cell-size was operated at $\rm V_{OV}$ = 1~V.} 
\label{fig:analysis}
\end{figure}

Events are selected by requiring an energy deposition in one crystal bar compatible with a MIP. Figure~\ref{fig:analysis} shows an example of MIP energy distribution measured in one bar. The energy corresponds to the average of the energies measured by the two individual SiPMs. Events are selected with energy between 0.80$\cdot$MPV, where MPV is the most probable value of a Landau function fitted to the MIP energy distribution, and a cut-off on the maximum amplitude to remove saturated signals.
An additional selection is applied on the impact point position of the particles by requiring coincidence with a MIP event in one central bar of the upstream reference module. This requirement has the effect of reducing the beam spot size to a few millimeters along the longitudinal axis of the bar under test.

In BTL, the time of arrival of a MIP in a single bar will be computed from the average of the two times, $t_L$ and $t_R$, measured at each bar end: $t_{bar} = 0.5 \cdot (t_L + t_R)$.
In this work, the bar time resolution is estimated as half of the spread of the difference $t_{diff} = t_L-t_R$:

\begin{equation}
    \sigma_{t_{bar}} = \frac{1}{2} \sigma_{t_{diff}}
    \label{eq:tDiff}
\end{equation}

which, in the absence of correlated uncertainties between the two time measurements $t_L$ and $t_R$ and for a fixed impact point position along the bar, is equivalent to the resolution of the average time $t_{bar}$. This method, already adopted in our previous work \cite{BTL_TB_paper_2021}, was verified again to yield a bar time resolution comparable to the one obtained from $t_{bar}$ on a subset of the current data set, where a micro-channel plate (MCP) detector located in front of the module was used as an external time reference.
Amplitude walk corrections are derived empirically by studying the $t_{diff}$ dependence on the ratio of the energies measured at the two bar ends. An example of time dependence on the energy ratio is reported in figure~\ref{fig:analysis}. The fitted function is used to correct the raw $t_{diff}$ in each event.

\begin{figure}[!tbp]
    \centering
    \includegraphics[width=0.49\linewidth]{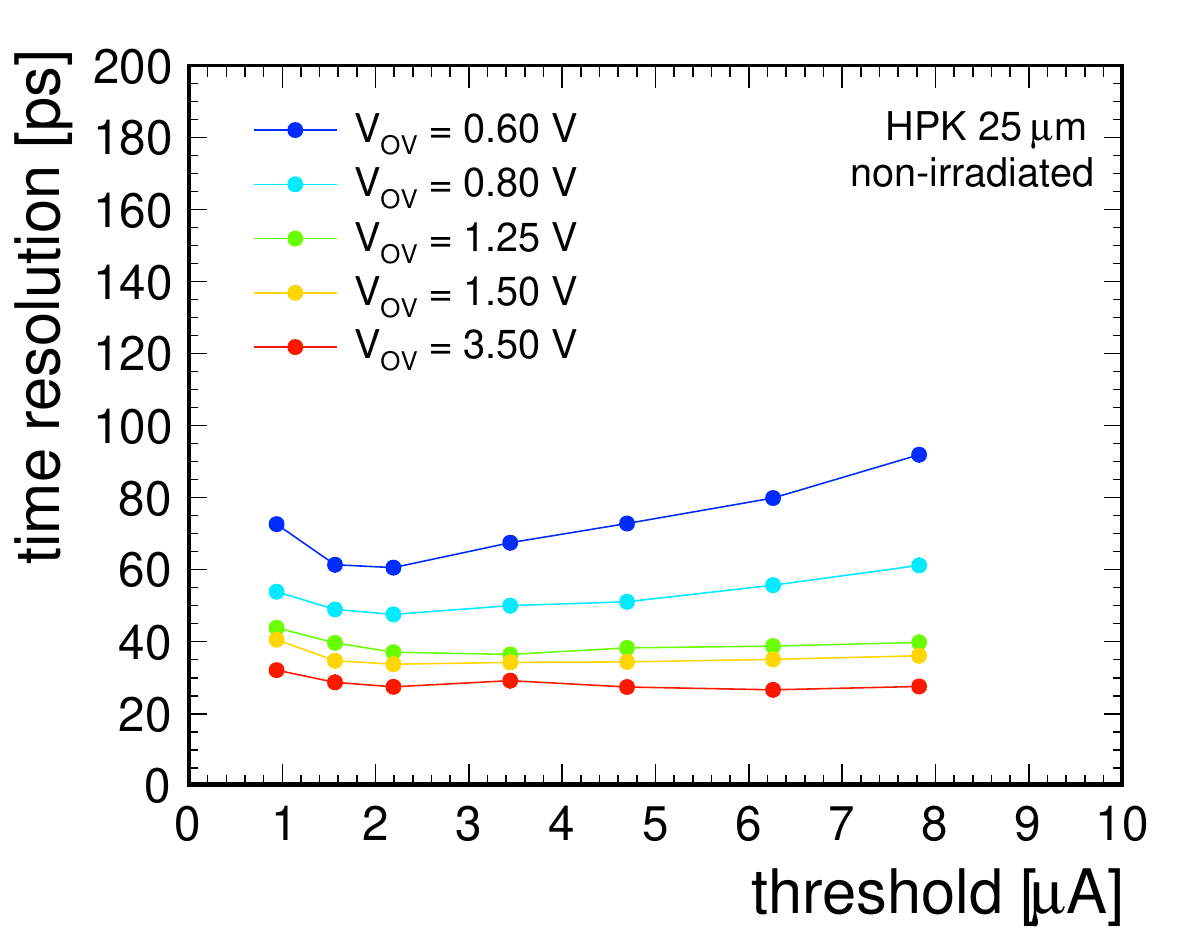}
    \includegraphics[width=0.49\linewidth]{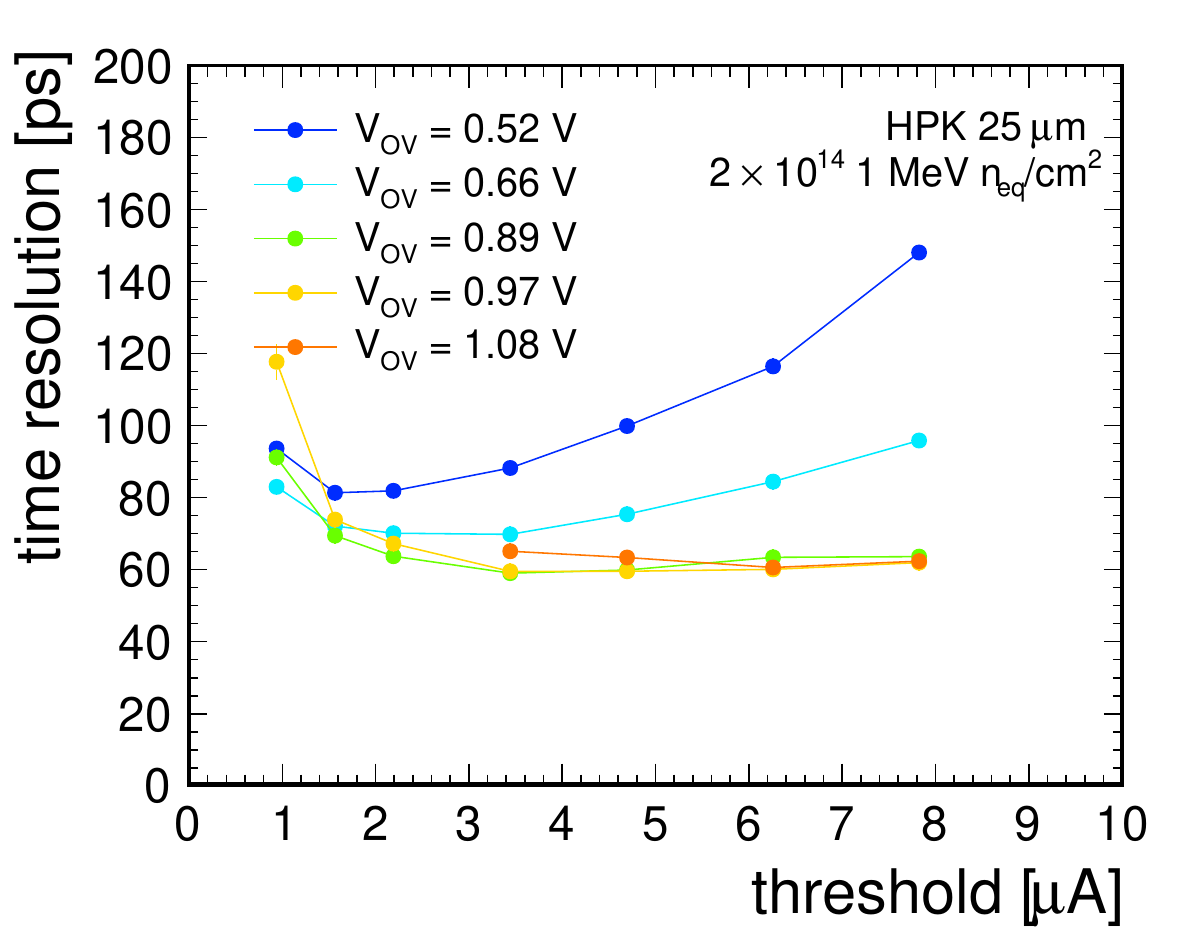}
    \caption{Time resolution as a function of the leading edge discrimination threshold for different values of the SiPM over-voltage. Left: T2 module with non-irradiated SiPMs. Right: T2 module with SiPMs irradiated to $2\times10^{14}$ \neut ~ operated at a temperature of $-35^\circ$C to emulate the DCR level of end-of-operation conditions.}
    \label{fig:timeResolution_vs_threshold}
\end{figure}

The leading edge discriminator features a current comparator for the pulses. For each configuration tested, a scan of the discriminator threshold is performed to find the one providing the best time resolution. 
Figure~\ref{fig:timeResolution_vs_threshold}(left) shows an example of time resolution measured for one bar in a module with non-irradiated SiPMs as a function of the leading edge discrimination threshold for various over-voltages. The set of thresholds used corresponds, in terms of single photoelectron amplitude, to a range from approximately 1 photoelectron (for 1$~\rm \mu A$, with a SiPM having a cell size of 25 $\mu$m operating at $V_{OV}$=3.5~V) to 32 photoelectrons (for 8$~\rm \mu A$, with the same SiPM operating at $V_{OV}$=1.0~V).

The time resolution is the result of two main contributions: one from stochastic fluctuations in the time of arrival of the photons, which increases as a function of the threshold, and one from the electronics noise, which decreases with increasing threshold. The typical optimal threshold is around 2-3~$\mu$A. For irradiated devices, an additional contribution to the time resolution is given by the DCR noise: the threshold needs to be further adjusted with increasing over-voltage (i.e. increasing DCR), as shown in figure~\ref{fig:timeResolution_vs_threshold}(right).

\section{Time resolution of modules with different cell-size SiPMs}
\label{sec:cell-sizes}

\begin{figure}[!tbp]
    \centering
    \includegraphics[width=0.49\linewidth]{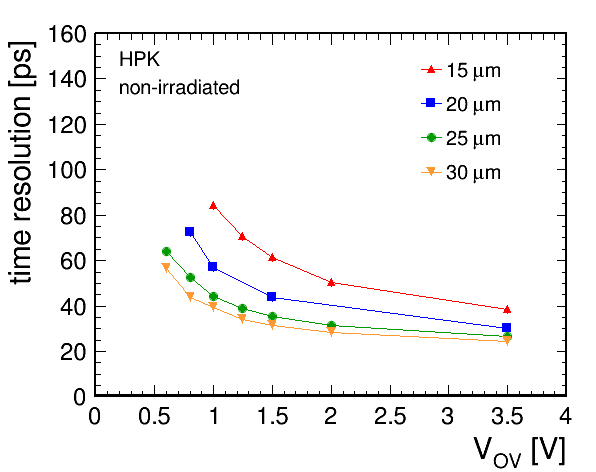}
    \includegraphics[width=0.49\linewidth] {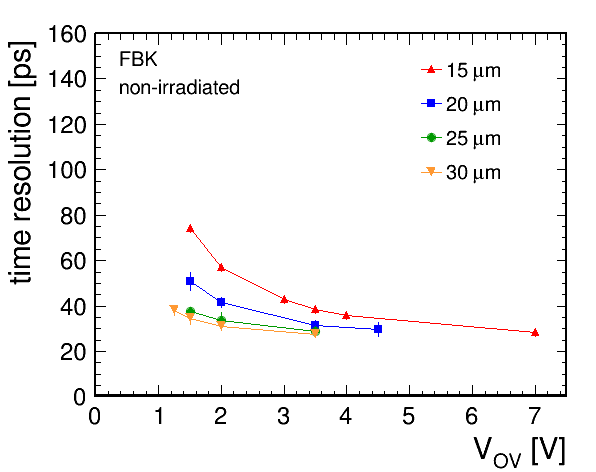}
    \caption{Time resolution as a function of the SiPM over-voltage for T2 modules with SiPMs of different cell-size. Left: modules with SiPMs from HPK. Right: modules with SiPMs from FBK.}
    \label{fig:timeResolution_nonIrradiated}
\end{figure}

The time resolution for T2 modules with non-irradiated SiPMs of different cell-sizes (15, 20, 25 and 30 $\mu$m) and from both HPK and FBK was measured as a function of the SiPM over-voltage. Each module under test was rotated by 52$^{\circ}$ to the beam direction to reproduce the most probable value of the MIP energy deposition expected in BTL (4.2~MeV, assuming the detector layout described in the MTD TDR~\cite{CMS_MTD_TDR}).
The results, averaged over several bars in each module, are reported in figure~\ref{fig:timeResolution_nonIrradiated}(left) for modules with SiPMs from HPK and in figure~\ref{fig:timeResolution_nonIrradiated}(right) for modules with SiPMs from FBK.
As expected, due to the higher PDE and higher gain,
SiPMs with larger cell-size provide better time resolution, reaching about 25-30~ps for an over-voltage of 3.5~V, which will be the typical working point for the BTL detector at the beginning of operation. A comparable performance is observed for both HPK and FBK SiPMs.

\begin{figure}[!t]
    \centering
    \includegraphics[width=0.49\linewidth]{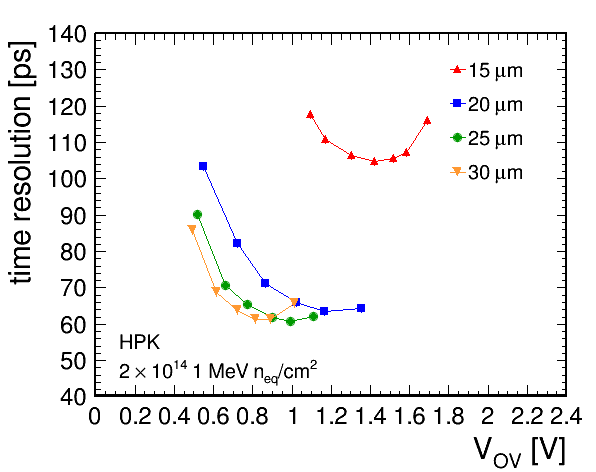}
    \includegraphics[width=0.49\linewidth]{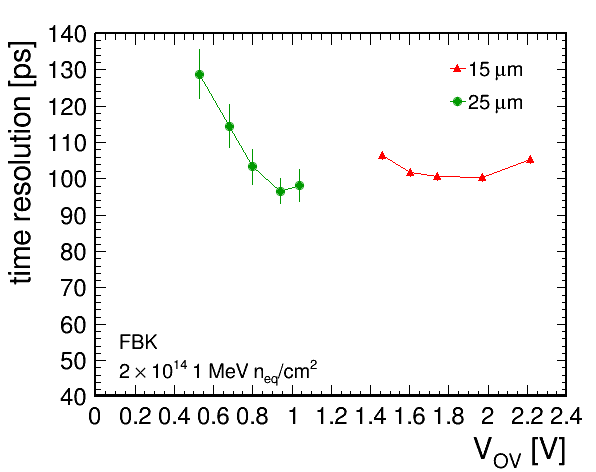}
    \caption{Time resolution as a function of the SiPM over-voltage for modules with SiPMs of different cell-sizes irradiated to 2$\times$10$^{14}$ \neut. Left: modules with SiPMs from HPK. Right: modules with SiPMs from FBK.}
    \label{fig:timeResolution_irradiated}
\end{figure}

The time resolution measured on modules with irradiated SiPM arrays is reported in figure~\ref{fig:timeResolution_irradiated} (irradiated SiPMs with 20 and 30 $\rm \mu m$ cell size from FBK were not available). SiPMs were irradiated to 2$\times$10$^{14}$ \neut~and operated at a temperature of $-35^\circ$C to reproduce the level of DCR expected at the end of the BTL operation. The optimal operating SiPM over-voltage is lower than that of modules with non-irradiated SiPMs, typically in the range between 0.8 and 1.4~V depending on the cells size. Operation at high over-voltage becomes prohibitive because of the steep and non-linear increase of the DCR with over-voltage (see figure~\ref{fig:dcr-curve}) which leads both to a large power dissipation that would lead to SiPM self-heating effects and to a degradation of the time resolution due to the contribution of the DCR term.

\begin{figure}[!tbp]
    \centering
    \includegraphics[width=0.49\linewidth]{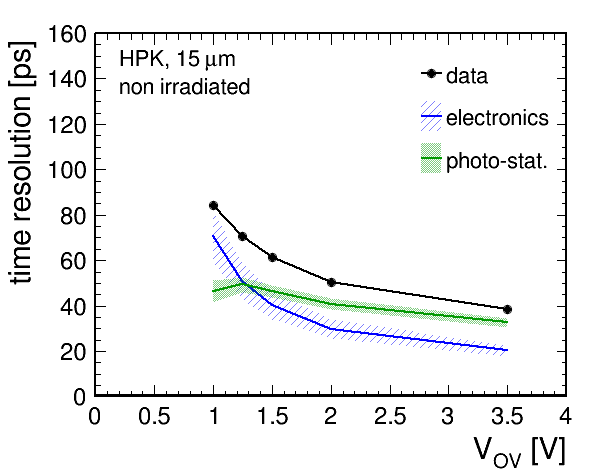}
    \includegraphics[width=0.49\linewidth]{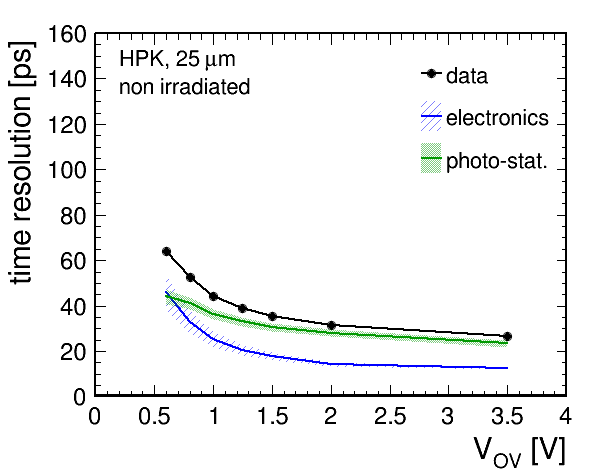}
    \includegraphics[width=0.49\linewidth]{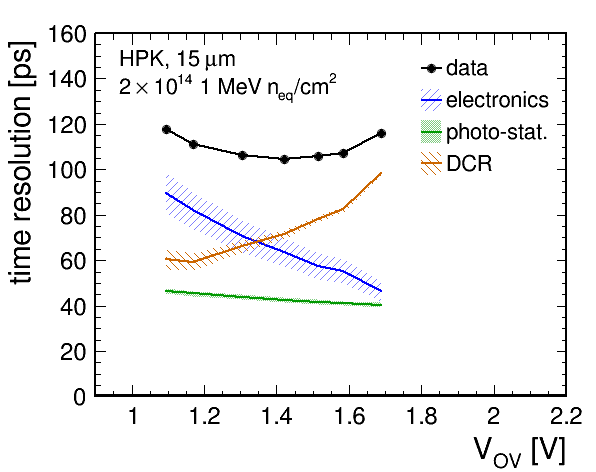}
    \includegraphics[width=0.49\linewidth]{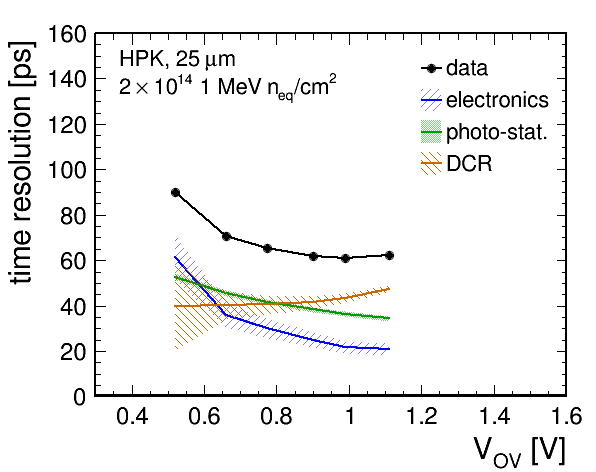}
    \caption{Time resolution as a function of over-voltage for T2 modules made with HPK SiPMs of 15~$\mu$m (left) and 25~$\mu$m (right) cell-size, both non-irradiated (top row) and irradiated to 2$\times$10$^{14}$ \neut. The time resolution measured with beam data is shown with black dots, the main contributions to the time resolution are shown by the colored lines: electronics (blue), the photo-statistics (green), and DCR (orange).}
    \label{fig:timeResolution_components}
\end{figure}

The superior performance of HPK SiPMs with larger cell-sizes can be understood by studying the different terms contributing to the time resolution. As discussed in section ~\ref{sec:drivers}, the main contributions are the photo-statistics component, the electronics noise and, for irradiated SiPMs, the DCR noise. 
Figure~\ref{fig:timeResolution_components} shows a quantitative example of the factorization of these different components for modules with 15~$\mu$m SiPMs and 25~$\mu$m cell size SiPMs from HPK.

\begin{figure}[!tbp]
    \centering
    \includegraphics[width=0.49\linewidth]{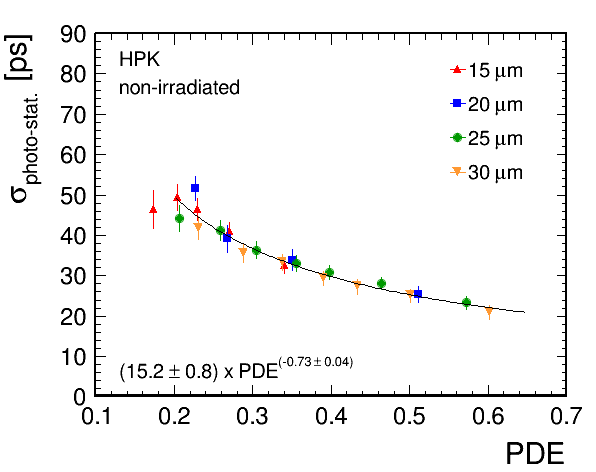}
    \includegraphics[width=0.49\linewidth]{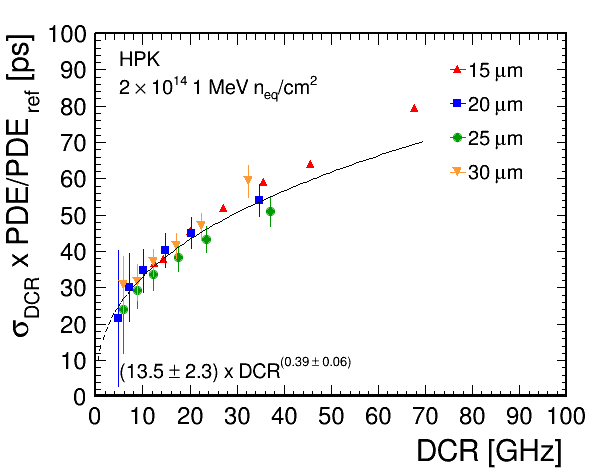}
    \caption{The photo-statistics component of the time resolution as a function of the SiPM PDE (left) 
    and DCR component as a function of the SiPM DCR (right) for T2 modules with non-irradiated SiPMs of different cell-sizes. The lines represent fits to data with a power law function in the form of $p_0 \times PDE^{\alpha}$ and $p_1 \times DCR^{\beta}$, respectively. $\rm PDE_{ref}$ corresponds to PDE of 25~$\mu m$ SiPMs from \vendorH~ at 1~V OV.}
    \label{fig:stochastic_vs_pde}
\end{figure}

\noindent
The electronics contribution is parameterized using eq.\ref{eq:ele} where the slope $dI/dt$ is estimated from the derivative of the average pulse shape as seen by the discriminator through a dedicated threshold scan and performing a linear fit of the pulse rising edge around the leading edge discrimination threshold. The uncertainty on the slope estimation, arising mainly from the variation of the range of the linear fit and differences between channels, is about 10\% and is reflected in the error band on the noise contribution of figure~\ref{fig:timeResolution_components}.
For modules with non-irradiated SiPMs, the photo-statistics term is estimated by subtracting in quadrature, at each over-voltage, the electronic contribution from the total measured time resolution. The behaviour of the photo-statistics component as a function of the PDE is shown in figure~\ref{fig:stochastic_vs_pde}~(left), where a global fit to data in the form $\sigma_{phot} = p_0 \times PDE^{\alpha}$ is superimposed yielding $\alpha = -0.73 \pm 0.04$. The slight departure from the prediction of the model of Section \ref{sec:drivers} is ascribed to threshold effects. For a fixed discrimination threshold (in units of single photoelectron amplitude) the time jitter due to photostatistcs is expected to scale as in Section~\ref{sec:drivers}, while if the effective discrimination threshold level is changed by selecting the optimal one at each over-voltage, as in the experimental results, this will also affect the time jitter due to photostatistics leading to a less trivial power law coefficient.
For irradiated SiPMs, the photo-statistics component reported in figure~\ref{fig:timeResolution_components} is estimated from the parametrization derived on non-irradiated modules and accounting for the PDE losses observed on irradiated devices.
The DCR component in figure~\ref{fig:timeResolution_components} is obtained by subtracting in quadrature the electronics and photo-statistics terms from the measured value of the total time resolution.
As shown in figure~\ref{fig:stochastic_vs_pde} (right), the scaling of the DCR component as a function of the DCR is close to the expectation (see eq.~\ref{eq:dcr}) with a coefficient $\beta = 0.39\pm 0.06$.

Close to the optimal operating over-voltage, larger cell-size SiPMs show a reduced impact from all the three terms on the time resolution (figure~\ref{fig:timeResolution_components}). The photo-statistics component improves thanks to the PDE being higher, despite the lower operating voltage, which implies a larger number of detected photo-electrons. Since both DCR and PDE increase with the SiPM effective active area, for irradiated devices, despite the larger DCR due to a better fill factor and better depletion of the cell, larger cell-size SiPMs provide a net gain thanks to the increase in signal-over-noise.
The larger SiPM gain and PDE increase the signal slope, thus suppressing significantly the electronics noise term, which is relevant especially at low operating over-voltages.

The performance of 15~$\rm \mu m$ HPK SiPMs is significantly worse than other cell sizes from the same manufacturer. According to the manufacturer, for 20-30~$\rm \mu m$ SiPMs they adopted an improved design to provide a larger PDE. Direct measurements of the PDE in our test stands confirm that the PDE of these samples scales relative to 15 um cells more than expected from simple geometrical considerations.
Moreover, a reduction of the signal amplitude (gain $\times$ PDE) of about 30\% was observed on 15~$\mu$m SiPMs irradiated to a fluence of $2\times10^{14}$ \neut~, while for larger cell-sizes the estimated signal loss is only about 20\%.
Similarly, the marginal improvement in time resolution observed for \vendorF~ irradiated SiPMs when increasing the cell size from 15 to 25~$\rm \mu m$ has been attributed to a loss of PDE of about 20\% after irradiation which was observed only in 25~$\rm \mu m$ SiPMs and not in 15~$\rm \mu m$ SiPMs. 
While further studies can be performed to improve the understanding of radiation damage effects in the SiPMs, the drift in breakdown voltage after irradiation is an indication of structural damage to the SiPMs with effects on the electric field inside the cells. These effects can be different in the cell center and periphery and, therefore, depend on the cell size.

The 25~$\mu$m cell size SiPMs from HPK were proven to be the optimal choice for the BTL detector within the allowed power budget of 30~mW/SiPM~\cite{CMS_MTD_TDR}.

\section{Time resolution of modules with different geometries}
\label{sec:geometries}

\begin{figure}[!tbp]
    \centering
    \includegraphics[width=0.49\linewidth]{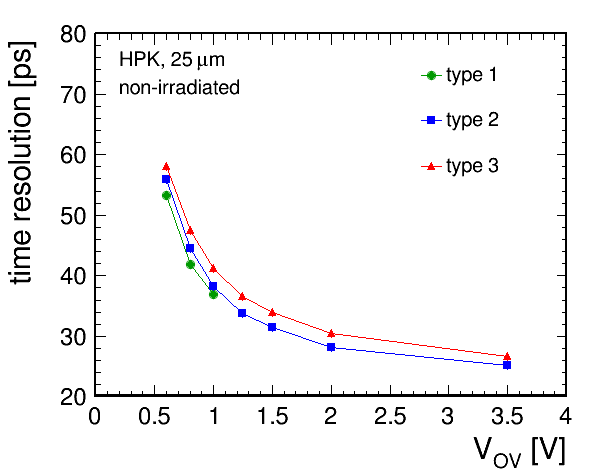}
    \includegraphics[width=0.49\linewidth]{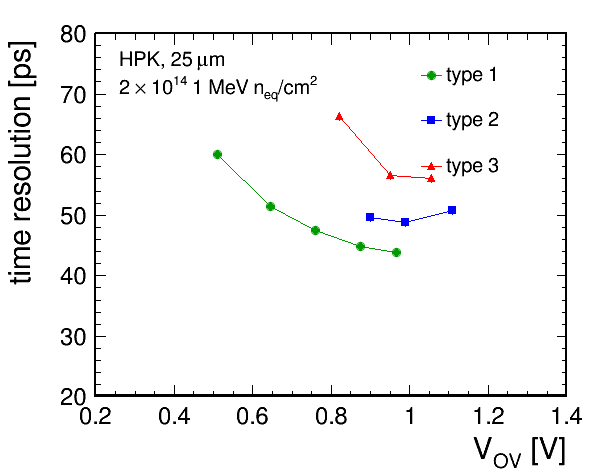}
    \caption{Time resolution as a function of the over-voltage for modules of different thicknesses (represented by green, blue, and red lines for T1, T2, and T3 modules, respectively).  Modules were rotated by an angle of 64$^{\circ}$ compared to the beam direction to emulate the energy deposition in the crystals expected in the outer pseudorapidity region ($|\eta| > 1.15$) of the BTL detector, as described in the MTD TDR~\cite{CMS_MTD_TDR}. Results are shown for both non irradiated (left) and irradiated to $2\times10^{14}$ \neut~ (right) modules.}
    \label{fig:types_timeResolution}
\end{figure}

The performance of T1, T2 and T3 modules (corresponding to crystal thicknesses of 3.75, 3.00 and 2.40~mm, respectively) was compared. The crystals were coupled to SiPM arrays matching the crystal geometry and with cell size of 25~$\mu$m.
Sensor modules made with both non-irradiated and irradiated SiPMs were characterized at a tilting angle of 64$^{\circ}$ to emulate the energy deposition in the crystals expected in the outer pseudorapidity region ($|\eta| > 1.15$) of the BTL detector, as described in the MTD TDR~\cite{CMS_MTD_TDR}.
The performance of the three modules was then compared in terms of time resolution as a function of the applied over-voltage as shown in figure~\ref{fig:types_timeResolution}. 

\begin{figure}[!htbp]
    \centering
    \includegraphics[width=0.49\linewidth]{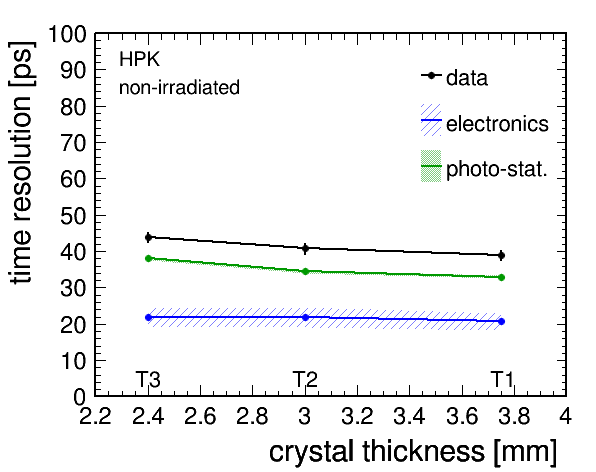}
    \includegraphics[width=0.49\linewidth]{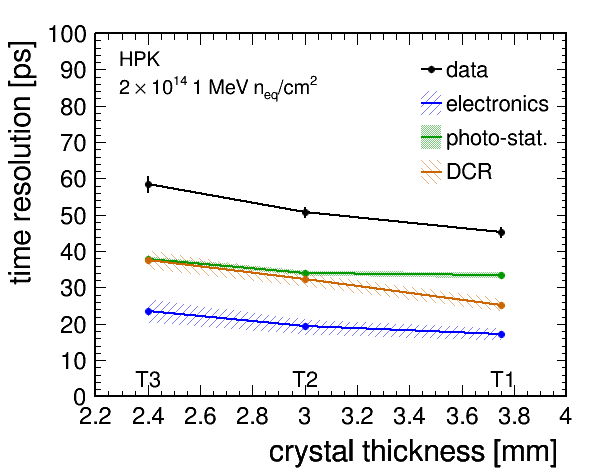}
    \caption{
    Time resolution of modules with non-irradiated (left) and irradiated (right) SiPMs as a function of the crystal thickness, i.e. for the T1 (3.75 mm), T2 (3.00 mm), and T3 (2.40 mm) modules. Results are shown for $\rm V_{OV}$ = 0.95~V. The terms contributing to the measured time resolution, photo-statistics (green), electronics (blue) and DCR (orange), are also reported.}
    \label{fig:res_vs_thickness}
\end{figure}

For non-irradiated modules the increase of crystal thickness is only marginally beneficial, because of two conflicting effects that affect the signal rising slope. Namely, the beneficial effect of a larger light signal, $N_{pe}$, due to a larger energy deposit combined with a better LCE (from a more favorable geometrical acceptance of optical photons in thicker crystals \cite{SIF_proceeding}) is counterbalanced by a slower SiPM response, $f_{SiPM}$, due to the larger capacitance in SiPMs with larger active area. 
This makes the electronic noise contribution to the time resolution almost independent of the thickness (figure~\ref{fig:res_vs_thickness}~(left)).
Conversely, for irradiated SiPMs the larger light signal in T1 sensor modules compared to thinner (T2, T3) modules brings the additional advantage of reducing the dominant DCR term with a sizable improvement of the time resolution. As shown in figure~\ref{fig:res_vs_thickness}~(right), the total time resolution measured on irradiated modules, being dominated by the photo-statistics and DCR terms improves approximately as the square root of the thickness ($\propto t^{-0.56}$).
This results from the combination of various effects. In particular, while the signal in photo-electrons increases with thickness because of the effects described above thicker crystals also require larger SiPM active areas and thus proportionally higher DCR.

\section{Time resolution of modules with SiPM irradiated at different fluences}
\label{sec:fluences}

A set of HPK SiPM arrays irradiated at intermediate neutron fluences, namely to $1\times10^{13}$~\neut~ and $1\times10^{14}$~\neut, were also used to build modules and assess the expected detector time resolution at an intermediate stage of its operation. This is of particular interest since some of the radiation damage effects (e.g. the increase of DCR) do not scale entirely linearly with the integrated neutron fluence.
All SiPM arrays were annealed with the same procedure described in section~\ref{sec:modules} and then operated at a temperature that reproduces the BTL expected DCR level after the corresponding fluence. 

\begin{figure}[!tbp]
    \centering  
    \includegraphics[width=0.49\linewidth]{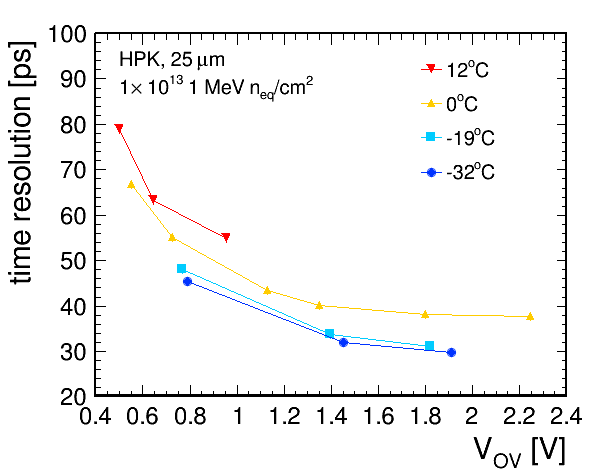} 
    \includegraphics[width=0.49\linewidth]{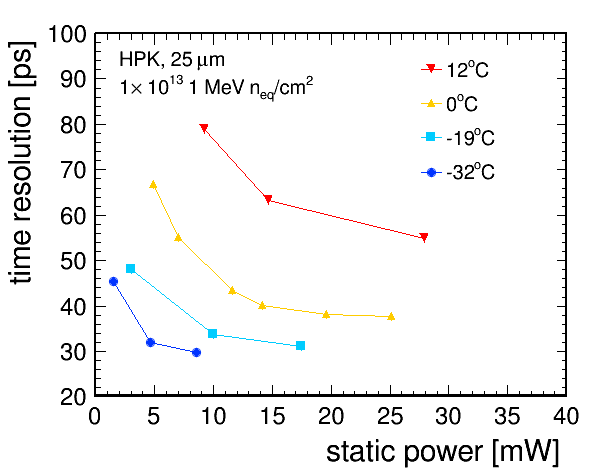} \\
    \includegraphics[width=0.49\linewidth]{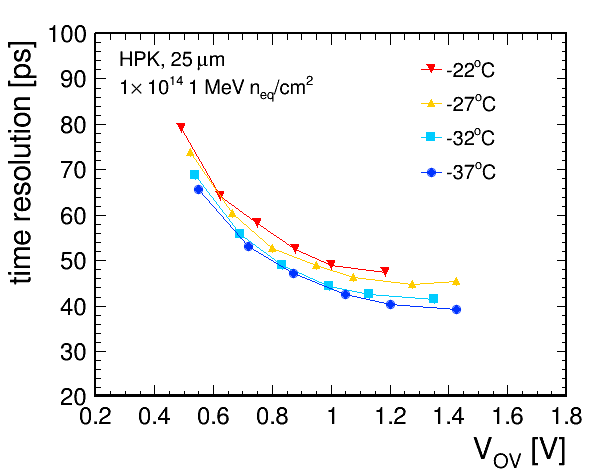}
    \includegraphics[width=0.49\linewidth]{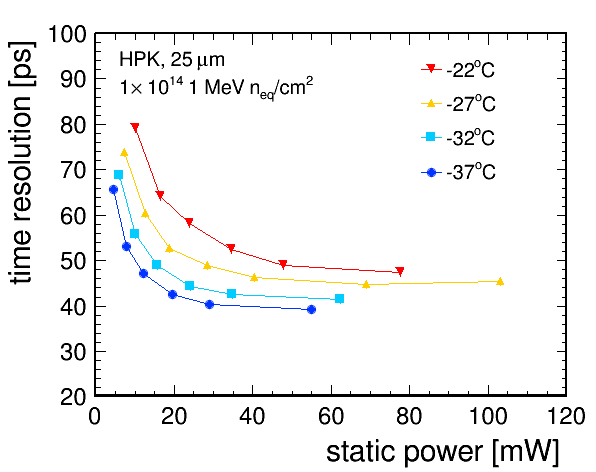}\\
    \includegraphics[width=0.49\linewidth]{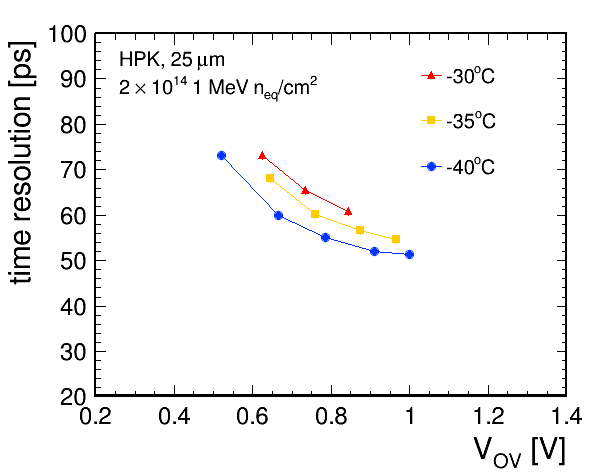}     
    \includegraphics[width=0.49\linewidth]{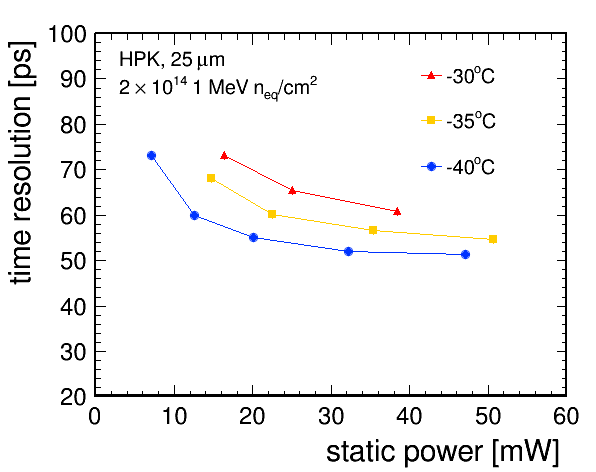} 
    \caption{Time resolution of three T1 modules (3.75~mm thick crystals) constructed using SiPM arrays with 25~$\mu$m and irradiated to different levels: $1\times10^{13}$~\neut~ (top), $1\times10^{14}$~\neut~ (middle), $2\times10^{14}$~\neut~ (bottom). All SiPMs underwent the same annealing. The time resolution is shown as a function of the over-voltage on the left column and as a function of static power of the SiPM on the right column. Different operating temperatures are compared around a temperature that yields a DCR level equivalent to the nominal BTL operating conditions (defined by irradiation level, annealing history and operating temperature).}
    \label{fig:tres_vs_fluences}
\end{figure}

The time resolution measured on these modules at various temperatures is shown in figure~\ref{fig:tres_vs_fluences}. While for non-irradiated modules the time resolution was demonstrated to be independent of the operating temperature, in irradiated SiPMs the increase of DCR by about a factor 2 every 10\degC is the leading factor that affects the change in time resolution.
For the modules with SiPMs irradiated to the highest fluence, the impact of temperature variation is the highest. In particular a $\pm 5$\degC change in temperature is translated into a about $\pm 5$~ps change in time resolution. The effect is smaller for lower irradiation levels since the relative contribution of the DCR term to the time resolution is smaller. In the BTL detector, slated to operate at $-45$\degC, the uniformity of temperature across all SiPMs is expected to be within a couple of degrees thus leading to a temperature related spread in time resolution across modules smaller than 5 ps.
    
The time resolution of the modules is also shown as a function of the static power per SiPM defined as the product of the SiPM operating bias voltage and its dark current. This comparison emphasizes the advantage of operating the SiPMs at a lower temperature since a lower static power corresponds to an overall smaller contribution of the SiPM to the power consumption of the full detector and to smaller self-heating effects on the SiPMs. In the BTL detector SiPMs will be operated with a power budget of about 30 mW. Within this constraint, the results show a time resolution of 30, 44, and 58~ps when operated in conditions representing the BTL irradiation and DCR levels after a half, five and ten years of HL-LHC operations, corresponding to a fluence of 0.1, 1.0, and $2.0\times10^{14}$~\neut, respectively.

\section{Conclusions}

The time resolution of a variety of BTL sensor module prototypes, consisting of LYSO:Ce crystal arrays coupled to different flavors of SiPM arrays, has been measured using minimum ionizing particles at the CERN and FNAL test beam facilities.
The comparison of SiPMs with different cell sizes ($15, 20, 25, 30 ~\mu$m) has been used to better understand the key parameters influencing the time resolution of the detector modules. While various effects come into play when changing the cell size it was observed that
the increase in gain and PDE provided by larger cell sizes provides a sizeable improvement in the time resolution, especially after irradiation. The three major contributions to the time resolution, namely, the photo-statistics, the electronics, and the DCR noise terms are overall smaller at the optimal over-voltage for larger cell SiPMs.

The results obtained show that the time resolution is well modeled as a function of few SiPM parameters (gain, PDE and DCR), as discussed in Section~\ref{sec:drivers}. However, effects of radiation damage which determine the evolution of such parameters with integrated neutron fluences vary depending on the cell size, on the SiPM manufacturer and on the specific SiPM technology. While an increase in the dark count rate is the dominant effect, observed losses in the PDE and gain provide indications of structural changes to the SiPMs.
A detailed study of the evolution of time resolution as a function of the operating temperature and of the neutron fluence to which SiPMs were exposed shows that performance degradation occurs smoothly with integrated fluence and is the combination of a quasi-linear increase in DCR with a radiation induced drop in signal amplitude. While the DCR can be reduced through annealing or by lowering the operating temperature, the latter effect cannot be mitigated in this way.
As an additional handle to improve the time resolution, the use of thicker LYSO:Ce crystals to boost the energy deposited by ionizing particles and consequently the number of photoelectrons was tested. The results confirm that the time resolution improves by a factor that scales approximately with the square root of the thickness.

In conclusion, the optimal choice for the BTL scope consists in $25~\mu$m SiPM cell size from \vendorH~. Modules of constant thickness at all pseudorapidities (T1 modules 3.75~mm thick) provide the best performance at an affordable cost increase of about 10\%. Such a sensor module configuration achieves a time resolution of about 25~ps with non-irradiated SiPMs and about 60~ps under the irradiation, annealing and temperature conditions representative of the end of the BTL detector operation.
Aside its immediate impact on the final choice of the MTD barrel detector technology, the present study outlines the key parameters that can be optimized for other specific applications of scintillator and SiPM based large area timing detectors aiming at few tens of ps resolution in either low or high radiation environments.

\acknowledgments

The authors are grateful to the technical experts of the CERN and FNAL beam line facilities for their invaluable help. We thank our colleagues Ryan Heller, Chris Madrid and Si Xie for their support in setting up the DAQ at FNAL. This project has received funding from the European Union’s Horizon Europe Research and Innovation programme under Grant Agreement No. 101057511 (EURO-LABS), MoST (China) under the National Key R\&D Program of China (No. 2022YFA1602100), the State Key Laboratory of Nuclear Physics and Technology, Peking University (No. NPT2023ZX01), and the Funda\c{c}\~ao para a Ci\^encia e a Tecnologia (FCT), Portugal.
We are also grateful to the FBK and HPK teams for their support in the long SiPM R\&D campaign which led to results presented in this paper.

\bibliography{mybibfile}

\end{document}

%% file: authors.tex
\author[27,a]{F.~Addesa,\note[a]{Now at  Paul Scherrer Institute PSI, Villigen, Switzerland}}  
\author[22]{T.~Anderson,}
\author[8,b]{P.~Barria,\note[b]{Now at IAPS - INAF, Rome, Italy}}
\author[8,9]{C.~Basile,}
\author[4]{A.~Benaglia,}
\author[4]{R.~Bertoni,} 
\author[28,c]{A.~Bethani,\note[c]{Now at Université Catholique de Louvain, Louvain-la-Neuve, Belgium}}
\author[8,19]{R.~Bianco,}
\author[26]{A.~Bornheim,}
\author[4,5,d]{G.~Boldrini,\note[d]{Now at Laboratoire Leprince-Ringuet, CNRS/IN2P3, Ecole Polytechnique, Institut Polytechnique de Paris, Palaiseau, France}}
\author[18]{A.~Boletti,}
\author[14,15]{A.~Bulla,}
\author[21]{M.~Campana,} 
\author[22]{B.~Cardwell,}
\author[4,5]{P.~Carniti,}
\author[4,5]{F.~Cetorelli,}
\author[4,5]{F.~De~Guio,}
\author[12,13]{K.~De~Leo,}
\author[8,9]{F.~De~Riggi,}
\author[21]{J.~Dervan,}
\author[22]{E.~Fernandez,} 
\author[16]{A.~Gaile,}
\author[18]{M.~Gallinaro,}
\author[4,5]{A.~Ghezzi,}
\author[4]{C.~Gotti,} 
\author[25]{R.~Goldouzian,} 
\author[4,5,e]{V.~Guglielmi,\note[e]{Now at Deutsches Elektronen-Synchrotron, Hamburg, Germany}} 
\author[25]{A.~Heering,}
\author[2]{Z.~Hu,}
\author[22]{M.~Jose,}
\author[25,f]{A.~Karneyeu,\note[f]{Also at Institute for Nuclear Research, Moscow, Russia}}
\author[21]{A.~Krishna,}
\author[28]{B.~Kronheim,} 
\author[20,g]{C.~Kuo,\note[g]{Now at Hamburg University, Hamburg, Germany}} 
\author[26]{A.~La~Torre,} 
\author[22,h]{A.~Li,\note[h]{Now at Institute of High Energy Physics (HEPHY), Austrian Academy of Sciences (ÖAW), Vienna, Austria}}
\author[8,9]{F.~Lombardi,}
\author[4,5]{M.T.~Lucchini,}
\author[4]{M.~Malberti,}
\author[1]{Y.~Mao,}
\author[21,i]{B.~Marzocchi,\note[i]{Now at University of Minnesota, Minneapolis, USA}}
\author[10,11]{P.~Meridiani,}
\author[25,f]{Y.~Musienko,}
\author[22]{C.~Neu,} 
\author[8,9]{G.~Organtini,} 
\author[21]{T.~Orimoto,} 
\author[28]{C.~Palmer,} 
\author[4,5]{M.~Paganoni,} 
\author[4,5]{S.~Palluotto,}
\author[8,9]{R.~Paramatti,}
\author[4]{G.~Pessina,}
\author[22]{R.~T.~Perelman,}
\author[8,9]{C.~Quaranta,}
\author[4]{N.~Redaelli,}
\author[4,5]{S.~Redaelli,}
\author[17]{G.~Reales~Gutiérrez,}
\author[4,5,l]{S.~Ronchi,\note[l]{Now at IRFU, CEA, Université Paris-Saclay, Gif-sur-Yvette, France}}
\author[8,9]{F.~Santanastasio,}
\author[23]{I.~Schmidt,}
\author[18]{J.~C.~Silva,}
\author[24]{G.~Sorrentino,}
\author[1]{X.~Sun,}
\author[4,5]{T.~Tabarelli~de~Fatis,}
\author[6,7]{M.~Tosi,}
\author[3]{B.~Ujvari,}
\author[18]{J.~Varela,}
\author[1]{J.~Wang,}
\author[25]{M.~Wayne,}
\author[22]{S.~White,}
\author[1,4,5,m]{L.~Zhang,\note[m]{Now at University of Maryland, College Park, USA}} 
\author[1]{M.~Zhang}

%% file: address.tex
\affiliation[1]{Department of Physics and State Key Laboratory of Nuclear Physics and Technology, Peking University, Beijing, China}
\affiliation[2]{Tsinghua University, Beijing, China}
\affiliation[3]{Institute of Physics, University of Debrecen, Debrecen, Hungary}
\affiliation[4]{INFN Sezione di Milano-Bicocca, Milano, Italy}
\affiliation[5]{Università di Milano-Bicocca, Milano, Italy}
\affiliation[6]{INFN Sezione di Padova, Padova, Italy}
\affiliation[7]{Università di Padova, Padova, Italy}
\affiliation[8]{INFN Sezione di Roma, Rome, Italy}
\affiliation[9]{Sapienza Università di Roma, Rome, Italy}
\affiliation[10]{INFN Sezione di Torino, Torino, Italy}
\affiliation[11]{Università di Torino, Torino, Italy}
\affiliation[12]{INFN Sezione di Trieste, Trieste, Italy}
\affiliation[13]{Università di Trieste, Trieste, Italy}
\affiliation[14]{INFN Sezione di Cagliari, Cagliari, Italy}
\affiliation[15]{Università degli Studi di Cagliari, Cagliari, Italy}
\affiliation[16]{Riga Technical University,  Riga, Latvia}
\affiliation[17]{Faculty of Mechanical, Maritime, and Materials Engineering, Delft University of Technology, Delft, The Netherlands}
\affiliation[18]{Laboratório de Instrumentação e Física Experimental de Partículas, Lisboa, Portugal}
\affiliation[19]{CERN, European Organization for Nuclear Research, Geneva, Switzerland}
\affiliation[20]{ National Taiwan University (NTU), Taipei, Taiwan}
\affiliation[21]{Northeastern University, Boston, USA}
\affiliation[22]{University of Virginia, Charlottesville, USA}
\affiliation[23]{The University of Iowa, Iowa City, USA}
\affiliation[24]{Kansas State University, Manhattan, USA}
\affiliation[25]{University of Notre Dame, Notre Dame, USA}
\affiliation[26]{California Institute of Technology, Pasadena, USA}
\affiliation[27]{Princeton University, Princeton, USA}
\affiliation[28]{University of Maryland, College Park, USA}